\documentclass[12pt, draftcls, onecolumn]{IEEEtran}

\ifCLASSINFOpdf
  \usepackage[pdftex]{graphicx}
  \DeclareGraphicsExtensions{.pdf,.jpeg,.png}
\else
  \usepackage[dvips]{graphicx}
\fi
\usepackage{subfigure}
\usepackage{epstopdf}
\usepackage[cmex10]{amsmath}
\usepackage{amssymb}
\usepackage{wasysym}
\usepackage{algorithm}
\usepackage{algorithmic}
\usepackage{array}
\usepackage{cite}
\usepackage{color}
\usepackage{url}
\usepackage{bm}
\usepackage{enumerate}
\usepackage{stfloats}
\usepackage{amsmath}
\usepackage{amsthm,amsmath}
\usepackage{mathrsfs}
\usepackage{epsfig,latexsym}
\usepackage{flushend}
\usepackage{cite}
\usepackage{hyperref}
\usepackage{diagbox} 
\usepackage{booktabs}

\begin{document}
%
\title{Hybrid Precoding for Mixture Use of Phase Shifters and Switches in mmWave Massive MIMO}
\author{Chenhao~Qi, Qiang~Liu, Xianghao~Yu and Geoffrey Ye Li
\thanks{Chenhao~Qi and Qiang Liu are with the School of Information Science and Engineering, Southeast University, Nanjing 210096, China (Email: qch@seu.edu.cn).}
\thanks{Xianghao Yu is with the Department of Electronic and Computer Engineering, The Hong Kong University of Science and Technology, Hong Kong 999077, China (Email: eexyu@ust.hk).}
\thanks{Geoffrey Ye Li is with the Department of Electrical and Electronic Engineering, Imperial College London, London SW7 2AZ, UK (Email: geoffrey.li@imperial.ac.uk).}
}

\markboth{Draft}
{Shell \MakeLowercase{\textit{et al.}}: Bare Demo of IEEEtran.cls for Journals}

\maketitle

\begin{abstract}
A variable-phase-shifter (VPS) architecture with hybrid precoding for mixture use of phase shifters and switches, is proposed for millimeter wave massive multiple-input multiple-output communications. For the VPS architecture, a hybrid precoding design (HPD) scheme, called VPS-HPD, is proposed to optimize the phases according to the channel state information by alternately optimizing the analog precoder and digital precoder. To reduce the computational complexity of the VPS-HPD scheme, a low-complexity HPD scheme for the VPS architecture (VPS-LC-HPD) including alternating optimization in three stages is then proposed, where each stage has a closed-form solution and can be efficiently implemented. To reduce the hardware complexity introduced by the large number of switches, we consider a group-connected VPS architecture and propose a HPD scheme, where the HPD problem is divided into multiple independent subproblems with each subproblem flexibly solved by the VPS-HPD or VPS-LC-HPD scheme. Simulation results verify the effectiveness of the propose schemes and show that the proposed schemes can achieve satisfactory spectral efficiency performance with reduced computational complexity or hardware complexity.
\end{abstract}



\section{Introduction}
Millimeter wave (mmWave) communications are capable of providing ultra-high-speed data rate owing to the abundant mmWave frequency resources~\cite{heath2016overview,SciChinaChenhao2021,CM2019HardwareConstrainedmmWave,RI2020Ta}. However, uplifting the carrier frequency to such high frequency band inevitably brings several challenges such as a significant path loss and unfavorable atmospheric absorptions~\cite{rappaport2013millimeter}, making it difficult for the mmWave signal to propagate in a long distance. To address these drawbacks, massive multiple-input multiple-output (MIMO) is introduced for mmWave communications to synthesize beams with high directivity and large beam gain~\cite{AsurveyICST2018}.

When increasing the operating frequency from sub-6 GHz to mmWave bands, the hardware cost of fully digital precoding becomes a bottleneck. To approach the achievable rate performance of the fully digital precoding meanwhile reducing the hardware cost, hybrid precoding and combining are proposed, where we can use much fewer radio frequency (RF) chains than antennas~\cite{ay2014spatially,Yli2016ee,AA2019DD,HybridICM2017,3DKoc}. According to the way that each RF chain connects to the antennas, the hybrid precoding can be categorized into the fully-connected~\cite{ch2017compress} and partially-connected architectures~\cite{sa2017par,Energy2016Gao}, where the latter uses fewer phase shifters than the former but with certain performance degradation. Nevertheless, either in the fully-connected or partially-connected architectures, the phase shifters play an important role~\cite{hcs,hbw}.

To reduce the hardware complexity and power consumption of phase shifters, the switches that only have on-off binary states are adopted for hybrid precoding and combining~\cite{rwh2016verse,Buz2018EE,LHY2020DH}. Some works even completely replace the phase shifters by the low-cost switches~\cite{zh2018on,no2019sw}. In~\cite{zh2018on}, two optimal signal-to-noise-ratio maximization algorithms are proposed to determine the binary states of the switches under the per-antenna power constraint and the total power constraint, which shows that full diversity gain and full array gain can be achieved with complexities only up to a polynomial order. In~\cite{no2019sw}, the problem of mutual information maximization is cast as a binary, rank-constrained quadratic maximization, which can be solved iteratively for each column of the analog precoder and approximated by a set of sequential convex programming. In~\cite{hong2018trans}, the switches are used for antenna selection to achieve further performance improvement. Based on the low-complexity beamformers derived with close-form expressions, quantized hybrid beamforming and channel estimation are developed~\cite{pay2018low}, while the achievable sum-rates is analyzed for frequency-selective channels~\cite{pay2020hy}. From an energy efficiency (EE) perspective~\cite{pay2019pssw}, the closed-form expressions are provided to compare several promising hybrid beamforming architectures and the optimal numbers of antennas maximizing the EE are obtained, where significantly higher EE can be achieved by the combination of phase shifters and switches than the conventional phase shifter-only architectures. In~\cite{HybridCombiningPhaseShiftersTVT2020}, an architecture using switches to select antenna subsets and using constant phase shifters to control the phases of signals in the RF circuit is proposed, where three low-complexity algorithms is developed for per-RF chain antenna subset selection. In~\cite{yu2019hard}, a fixed phase shifter (FPS) architecture, with the mixture use of the phase shifter and switch networks, performs precoding based on alternating minimization (AltMin).

In this paper, by combining the advantages of high resolution of phase shifters and low cost of switches, we consider the hybrid precoding design (HPD) for mixture use of phase shifters and switches in mmWave massive MIMO. The main contributions are summarized as follows.
\begin{itemize}
	 \item Different from the existing FPS architecture where the phases are fixed and independent of channel state information (CSI), we propose a variable-phase-shifter (VPS) architecture whose phases are variable and can be optimized according to the CSI subject to the hardware constraints.

\item Based on the VPS architecture, a HPD scheme called VPS-HPD, is proposed to optimize the phases according to the CSI. Specifically, we alternately optimize the analog precoder and the digital precoder, where the analog precoder optimization is converted into several subproblems and each subproblem further includes the alternating optimization of the phase matrix and switch matrix.

\item To reduce the computational complexity of the VPS-HPD scheme, we propose a low-complexity HPD scheme for the VPS architecture (VPS-LC-HPD) that does not need Riemannian manifold optimization and exhaustive search. The VPS-LC-HPD scheme includes alternating optimization in three stages, where each stage has a closed-form solution and can be efficiently implemented.

\item To reduce the hardware complexity introduced by large number of switches, we consider a group-connected VPS (GC-VPS) architecture, where each phase shifter is only connected to a group of antennas instead of all antennas through switches. Then we propose a HPD scheme for the GC-VPS architecture, where the HPD problem is divided into multiple independent subproblems with each subproblem flexibly solved by the VPS-HPD or VPS-LC-HPD scheme.

\end{itemize}

The remainder of this paper is organized as follows. In Section~\ref{sec.Systemfirst}, we introduce the system model. In Section~\ref{sec.ProblemFormulation}, we propose the VPS architecture. Then in Section~\ref{sec.HybridPrecoding}, we propose a VPS-HPD scheme. To reduce the computational complexity of VPS-HPD, a VPS-LC-HPD scheme is proposed in Section~\ref{sec.lowcomplexity}. To reduce the hardware complexity, the GC-VPS with the corresponding HPD scheme is proposed in Section~\ref{sec.group}. Simulation results are provided in Section~\ref{sec.Simulation}. Finally, Section~\ref{sec.conclusion} concludes this paper.

\textit{Notations}: Symbols for vectors (lower case) and matrices (upper case) are in boldface. For a vector $\boldsymbol{a}$, $[\boldsymbol{a}]_{m}$ denotes its $m$th entry. For a matrix $\boldsymbol{A}$, $[\boldsymbol{A}]_{m,:}$, $[\boldsymbol{A}]_{:,n}$, $[\boldsymbol{A}]_{m,n}$, ${\boldsymbol A}^T$, ${\boldsymbol A}^{-1}$, ${\boldsymbol A}^H$ and $\|\boldsymbol A\|^2_F$ denote the $m$th row, the $n$th column, the entry on the $m$th row and $n$th column, the transpose, the inverse, the conjugate transpose (Hermitian), and Frobenius norm, respectively. $\boldsymbol{I}_{L}$ denotes an $L \times L$ identity matrix. The symbols ${\angle}(\cdot)$, ${\mathbb{E}}(\cdot)$, ${\rm vec}(\cdot)$, ${\rm tr}(\cdot)$, ${\rm{Re}}(\cdot)$, $\mathcal{O}(\cdot)$ and $\mathcal{CN}({ m},\boldsymbol{R})$ denote the angle of a complex-valued number, the expectation, the vectorization, the trace, the real part of a complex-valued number, the order of complexity, and the complex Gaussian distribution with the mean of $m$ and the covariance matrix being $\boldsymbol{R}$, respectively. The symbols $\mathbb{C}$, $\mathbb{Z}$, and $\otimes$ denote the set of complex-valued numbers, the set of integers, and the Kronecker product, respectively.


\begin{figure*}[!t]
\centering
\includegraphics[width=165mm]{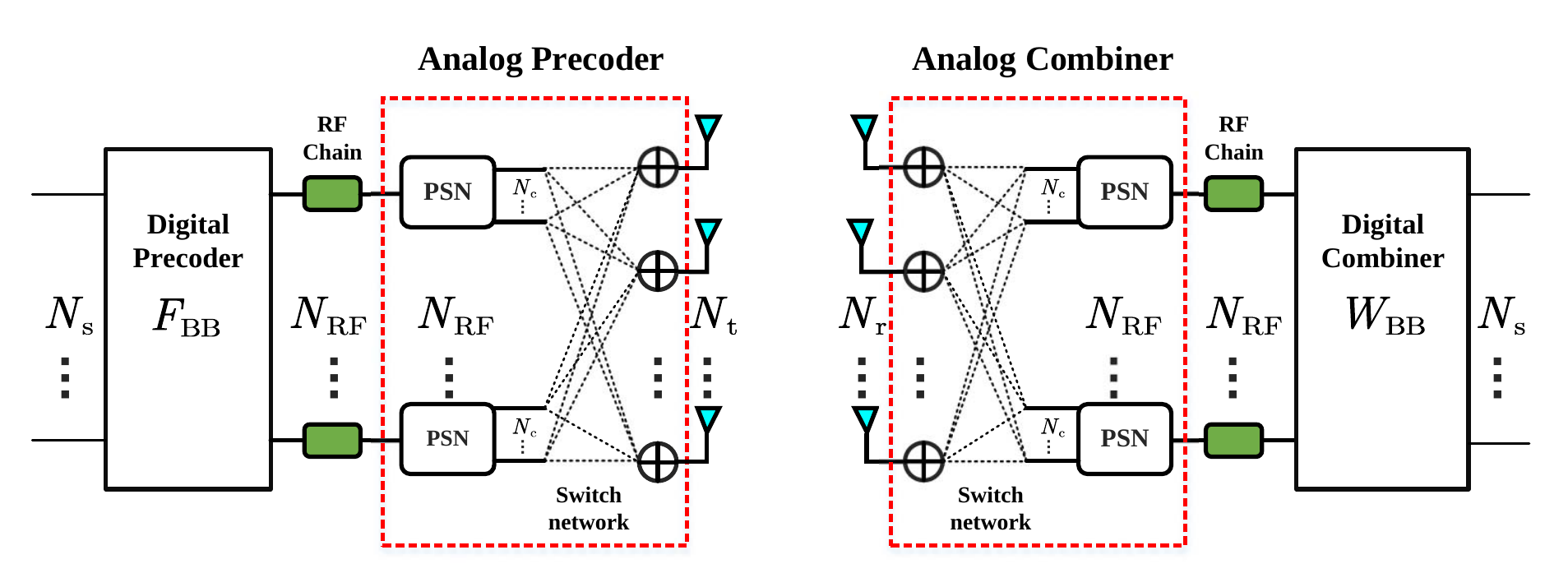}
\caption{A point-to-point mmWave massive MIMO system with the FPS/VPS architecture.}
\label{fig:system}
\end{figure*}

\section{System Model}\label{sec.Systemfirst}
Consider a point-to-point mmWave massive MIMO system with $N_{\rm t}$ and $N_{\rm r}$ antennas at the transmitter and receiver, respectively. The transmitter and receiver employ the hybrid precoding and combining with $N_{\rm RF}$ RF chains, where $N_{\rm RF} \ll N_{\rm t}$ and $N_{\rm RF} \ll N_{\rm r}$. We assume that $N_{\rm s}$ independent data streams are transmitted in parallel, where $N_{\rm s} \leq N_{\rm RF} $. The hybrid precoder includes a digital precoder at baseband (BB) and an analog precoder in the RF domain, denoted as $\boldsymbol{F}_{\rm BB}\in \mathbb{C}^{N_{\rm RF} \times N_{\rm s}}$ and $\boldsymbol{F}_{\rm RF}\in \mathbb{C}^{N_{\rm t} \times N_{\rm RF}}$, respectively. We normalize the power gain of the hybrid precoder by setting $\|\boldsymbol{F}_{\rm RF}\boldsymbol{F}_{\rm BB}\|^2_F=N_{\rm s}$. Similarly, the hybrid combiner includes a digital combiner and an analog combiner, denoted as $\boldsymbol{W}_{\rm BB}\in \mathbb{C}^{N_{\rm RF}\times N_{\rm s}}$ and $\boldsymbol{W}_{\rm RF}\in \mathbb{C}^{N_{\rm r}\times N_{\rm RF}}$, respectively. Then the received signal after hybrid combining can be expressed as
\begin{equation}\label{system model}
\boldsymbol y=\sqrt{P}\boldsymbol{W}_{\rm BB}^H\boldsymbol{W}_{\rm RF}^H\boldsymbol{H}\boldsymbol{F}_{\rm RF}\boldsymbol{F}_{\rm BB}\boldsymbol{s}  + \boldsymbol{W}_{\rm BB}^H\boldsymbol{W}_{\rm RF}^H \boldsymbol{\eta},
\end{equation}
where $\boldsymbol{s}\in \mathbb{C}^{N_{\rm s}}$ is the transmit signal such that $\mathbb{E}(\boldsymbol{s}\boldsymbol{s}^{H})=\frac{1}{N_{\rm s}}\boldsymbol{I}_{N_{\rm s}}$, $P$ is the transmit power, and $\boldsymbol{\eta}\sim\mathcal{CN}(0,\sigma^2\boldsymbol{I}_{N_{\rm r}})$ is a noise term which obeys the complex Gaussian distribution with zero mean and variance of $\sigma^2$.

Using the widely-used Saleh-Valenzuela model, the channel matrix $\boldsymbol{H}$ in \eqref{system model} is defined as
\begin{equation}\label{channelmodel}
\boldsymbol{H}=\sqrt{\frac{N_{\rm t}N_{\rm r}}{L}} \sum_{l=1}^{L}\alpha_{l}\boldsymbol{a}(N_{\rm r},\phi_{l})\boldsymbol{a}^{H}(N_{\rm t},\theta_{l}),
\end{equation}
where $L$, $\alpha_{l}$, $\theta_{l} \in (0, 2\pi)$ and $\phi_{l} \in (0, 2\pi)$ are the number of channel paths, complex channel coefficient, physical angle-of-departure (AoD) and physical angle-of-arrival (AoA) of the $l$th path for $l=1,2,\ldots,L$, respectively. Due to the limited scattering of mmWave signal, the number of resolvable channel paths is much smaller than the number of transmit antennas, i.e., $L\ll N_{\rm t}$, where $L$ is typically 3 to 5~\cite{ma2020high,Alk2014channelesti}. The channel steering vector, as a function of antenna number $N$ and AoD or AoA $\varphi$, is defined as
\begin{equation}
  \boldsymbol{a}(N,\varphi)\triangleq\frac{1}{\sqrt{N}}{\big[1,e^{j\pi\sin \varphi},e^{j2\pi\sin \varphi},\ldots,e^{j(N-1)\pi\sin \varphi}\big]}^T
\end{equation}
for uniform linear arrays (ULAs).

\section{VPS architecture}\label{sec.ProblemFormulation}
For the fully-connected architecture where each RF chain connects to all the antennas via phase shifters, we need totally $N_{\rm t}N_{\rm RF}$ phase shifters. To reduce the hardware complexity, the partially-connected
architecture decreases the number of phase shifters to $N_{\rm t}$, where each RF chain only connects to a subset of the antennas. Since fewer phase shifters are used, there is some performance degradation comparing to the fully-connected one.

Note that the switches with on-off binary states are much cheaper and faster than the phase shifters~\cite{rwh2016verse}. Therefore, the phase shifters in the fully-connected architecture can be replaced by low-cost switches with some sacrifice of achievable rate performance~\cite{zh2018on}. To balance the hardware complexity and achievable rate performance, an architecture named FPS with mixture use of phase shifters and switches is proposed~\cite{yu2019hard}. The block diagram of the FPS architecture is shown in Fig.~\ref{fig:system}, where the analog precoder or combiner is formed by $N_{\rm RF}$ phase shifter networks (PSNs), a switch network and $N_{\rm t}$ signal adders. Each PSN includes $N_{\rm c}$ phase shifters and therefore the FPS architecture includes $N_{\rm c} N_{\rm RF}$ phase shifters in total. In Table~\ref{Table1}, we list different numbers of phase shifters required for the fully-connected, partially-connected and FPS architectures. Since we normally set $N_{\rm c} \ll  N_{\rm t}$, we may use much fewer phase shifters in the FPS architecture than that in the fully-connected architecture. The switch network of the FPS architecture includes $N_{\rm c}N_{\rm RF} N_{\rm t}$ switches, where each switch determines the on-off state of the link between a PSN output and a signal adder. Each signal adder adds at most $N_{\rm c}N_{\rm RF}$ different signals together.

For the FPS architecture, the phase of the $i$th phase shifter in each PSN is $2\pi (i-1)/N_{\rm c} $ for $i=1,2,\ldots,N_{\rm c}$. Note that the phases are determined by $N_{\rm c}$. Moreover, the phases are fixed and independent of mmWave CSI, which does not fully exploit the phase-space freedom of the phase shifters.

Different from the FPS architecture, in this work we will consider a VPS architecture, where the phases are optimized according to the mmWave CSI and subject to the hardware constraints. Note that the numbers of phase shifters and switches are the same in both the VPS and FPS architectures, as shown in Table~\ref{Table1}. Each RF chain in both the FPS and VPS architectures is connected to all the antennas via phase shifters and switches, which shares a similar feature as the fully-connected architecture. However, the VPS uses variable phase shifters while the FPS uses fixed ones. Due to the difference of these two architectures, their HPD schemes are completely different.

\begin{table}[!t]
\centering
\caption{Overhead comparison for different mmWave MIMO architectures.}
\label{Table1}
\begin{tabular}{p{3.55cm}p{3.82cm}p{3.26cm}}
\toprule
Different architectures     &   Number of phase shifters &   Number of switches \\
\midrule
Fully-connected            &  $N_{\rm t}N_{\rm RF}$  & 0   \\
Partially-connected        &  $N_{\rm t}$            & 0    \\
FPS / VPS              &  $N_{\rm c}N_{\rm RF}$     & $N_{\rm c}N_{\rm RF}N_{\rm t}$\\
GC-VPS              &  $N_{\rm c}N_{\rm RF}$     & $N_{\rm c}N_{\rm RF}N_{\rm t}/q$\\
\bottomrule
\end{tabular}
\end{table}

According to Fig.~\ref{fig:system}, $\boldsymbol{F}_{\rm RF}$ can be expressed as
\begin{equation}\label{newanalog}
\boldsymbol{F}_{\rm RF}=\boldsymbol{S}_{\rm t}\boldsymbol{P}_{\rm t},
\end{equation}
where $\boldsymbol{P}_{\rm t}\in \mathbb{C}^{N_{\rm c}N_{\rm RF} \times N_{\rm RF}}$ is a phase matrix determined by $N_{\rm RF}$ PSNs and $\boldsymbol{S}_{\rm t}\in \mathbb{Z}^{N_{\rm t} \times N_{\rm c}N_{\rm RF}}$ is a switch matrix determined by the switch network. Each entry of $\boldsymbol{S}_{\rm t}$ is selected from a binary set
\begin{equation}
  \mathcal{A}\triangleq \{0,1\}
\end{equation}
where 1 and 0 indicate the on and off states of a switch, respectively. In fact, $\boldsymbol{P}_{\rm t}$ is a generalized block diagonal matrix, where each column of $\boldsymbol{P}_{\rm t}$ has only $N_{\rm c}$ nonzero entries representing the values of $N_{\rm c}$ phase shifters. Note that the shifting time of phase shifters is on the order of microseconds and therefore is much smaller than the channel coherent time typically on the order of milliseconds. The entry on the $i$th column and $((i-1)N_{\rm c}+l)$th row of $\boldsymbol{P}_{\rm t}$ can be denoted as
\begin{align}\label{PhaseMatrix}
  [\boldsymbol{P}_{\rm t}]_{(i-1)N_{\rm c}+l,i}=\frac{1}{\sqrt{N_{\rm c}}}e^{j\theta_{li}}, 
  i=1,2,&\ldots,N_{\rm RF}, ~l=1,2,\ldots,N_{\rm c},
\end{align}
where $\theta_{li}$ is the phase of the $l$th phase shifter in the $i$th PSN. Suppose the resolution of the phase shifters is $b$ bits~\cite{CCJ2020Two}. Then all available phases form a set
\begin{equation}
  \mathcal{B} \triangleq\left\{\frac{2 \pi i}{2^b}|i=1,2,...,2^b\right\}
\end{equation}
and we have $\theta_{li} \in \mathcal{B}$.

Suppose we have already obtained a channel estimate of $\boldsymbol {H}$, denoted as $\widehat{\boldsymbol {H}}$, using existing channel estimation methods~\cite{mwy,Alk2014channelesti}. The singular value decomposition (SVD) of $\widehat{\boldsymbol {H}}$ can be expressed as
\begin{equation}\label{svd}
  \widehat{\boldsymbol {H}} = \boldsymbol{V} \boldsymbol{}\Lambda \boldsymbol{U}^H
\end{equation}
where $\boldsymbol{U}$ is a unitary matrix with the dimension of $N_{\rm t}$ and $\boldsymbol{V} $ is a unitary matrix with the dimension of $N_{\rm r}$. The fully digital precoder $\boldsymbol{F}_{\rm opt}$ is determined by selecting the first $N_{\rm s}$ columns from $\boldsymbol{U}$~\cite{ay2014spatially}, i.e.,
\begin{equation}\label{SVD-Fopt}
	\boldsymbol{F}_{\rm opt}=[\bm U]_{:,1:N_{\rm s}}.
\end{equation}
The fully digital combiner $\boldsymbol{W}_{\rm opt}$ is determined by selecting the first $N_{\rm s}$ columns from $\boldsymbol{V}$ as $\boldsymbol{W}_{\rm opt}=[\bm V]_{:,1:N_{\rm s}}$.

The design of $\boldsymbol{F}_{\rm RF}$ and $\boldsymbol{F}_{\rm BB}$ aims to approximate $\boldsymbol{F}_{\rm opt}$ by
minimizing their Euclidean distance, which can be written as
\vspace{-0.2cm}
\begin{subequations}\label{solveto}
\begin{align}
& \min_{\boldsymbol{S}_{\rm t},\boldsymbol{P}_{\rm t},\boldsymbol{F}_{\rm BB}} {\big\|\boldsymbol{F}_{\rm opt}-\boldsymbol{S}_{\rm t}\boldsymbol{P}_{\rm t}\boldsymbol{F}_{\rm BB}\big\|}^2_F\\
&~~~~{\rm s.t.}  \ \ \ \ \angle\big( [\boldsymbol{P}_{\rm t}]_{i,l} \big)\in \mathcal{B},~\Big|[\boldsymbol{P}_{\rm t}]_{i,l}\Big|=\frac{1}{\sqrt{N_{\rm c}}},~\forall i,l, \label{constraint1}\\
& \ \ \ \ \ \ \ \ \ \ \ \ [\boldsymbol{S}_{\rm t}]_{m,n}\in\mathcal{A},~\forall m,n, \label{constraint2}\\
& \ \ \ \ \ \ \ \ \ \ \ \ {\|\boldsymbol{S}_{\rm t}\boldsymbol{P}_{\rm t}\boldsymbol{F}_{\rm BB}\|}^2_F=N_{\rm s}. \label{constraint3}
\end{align}
\end{subequations}
In fact, it has been shown that minimizing the objective function in \eqref{solveto} approximately leads to the maximization of the spectral efficiency or the sum-rate~\cite{Xianghao2016alter}. Note that the constraint in \eqref{constraint3} can be temporarily neglected. Then \eqref{solveto} can be rewritten as
\begin{subequations}\label{solveto2}
\begin{align}
& \min_{\boldsymbol{S}_{\rm t},\boldsymbol{P}_{\rm t},\boldsymbol{F}_{\rm BB}} {\big\|\boldsymbol{F}_{\rm opt}-\boldsymbol{S}_{\rm t}\boldsymbol{P}_{\rm t}\boldsymbol{F}_{\rm BB}\big\|}^2_F\\
&~~~~{\rm s.t.}  \ \ \ \ \angle\big( [\boldsymbol{P}_{\rm t}]_{i,l} \big)\in \mathcal{B},~\Big|[\boldsymbol{P}_{\rm t}]_{i,l}\Big|=\frac{1}{\sqrt{N_{\rm c}}},~\forall i,l, \label{constraint12}\\
& \ \ \ \ \ \ \ \ \ \ \ \ [\boldsymbol{S}_{\rm t}]_{m,n}\in\mathcal{A},~\forall m,n. \label{constraint22}
\end{align}
\end{subequations}
Once the solutions of \eqref{solveto2} is obtained, we can adjust $\boldsymbol{F}_{\rm BB}$ to satisfy \eqref{constraint3}~\cite{Xianghao2016alter}. Note that \eqref{solveto2} is a non-convex NP-hard optimization problem. In the following, we will elaborate on how to solve it.


\section{Hybrid Precoding Design}\label{sec.HybridPrecoding}
The optimization problem of \eqref{solveto2} involves three matrices $\boldsymbol{S}_{\rm t}$, $\boldsymbol{P}_{\rm t}$ and $\boldsymbol{F}_{\rm BB}$, which are coupled and difficult to handle. Therefore, we consider using alternating optimization to iteratively solve it. Specifically, we alternately optimize the analog precoder and the digital precoder, where the analog precoder optimization is converted into several subproblems and each subproblem further includes the alternating optimization of the phase matrix and switch matrix. In the following, we first determine $\boldsymbol{S}_{\rm t}$ and $\boldsymbol{P}_{\rm t}$ for given $\boldsymbol{F}_{\rm BB}$ in the first subsection, then determine $\boldsymbol{F}_{\rm BB}$ for given $\boldsymbol{S}_{\rm t}$ and $\boldsymbol{P}_{\rm t}$ in the second subsection, and finally analyze the convergence and computational complexity in the last subsection.

\subsection{Phase Matrix and Switch Matrix Optimization for Analog Precoder Design}\label{subsec.Matrixsolution}
Given $\boldsymbol{F}_{\rm BB}$, an estimate of $\boldsymbol{F}_{\rm RF}$, denoted as $\widehat{\boldsymbol{F}}_{\rm RF}$, can be expressed as
\begin{equation}\label{Analogm}
\widehat{\boldsymbol{F}}_{\rm RF}=\boldsymbol{F}_{\rm opt} \boldsymbol{F}_{\rm BB}^H (\boldsymbol{F}_{\rm BB} \boldsymbol{F}_{\rm BB}^H)^{-1}.
\end{equation}
Then we can determine $\boldsymbol{S}_{\rm t}$ and $\boldsymbol{P}_{\rm t}$ by the following optimization problem as
\begin{align}\label{Secondstep}
& \min_{\boldsymbol{S}_{\rm t}, \boldsymbol{P}_{\rm t}} \ {\big\|\widehat{\boldsymbol{F}}_{\rm RF}-\boldsymbol{S}_{\rm t}\boldsymbol{P}_{\rm t}\big\|}^2_F \notag\\
&~~{\rm s.t.} ~~ \ \angle\big( [\boldsymbol{P}_{\rm t}]_{i,l} \big)\in \mathcal{B},~\Big|[\boldsymbol{P}_{\rm t}]_{i,l}\Big|=\frac{1}{\sqrt{N_{\rm c}}},~\forall i,l, \notag\\
& \ \ \ \ \ \ \ \ \ [\boldsymbol{S}_{\rm t}]_{m,n}\in\mathcal{A},~\forall m,n.
\end{align}

We denote the nonzero entries of the $i$th column of $\boldsymbol{P}_{\rm t}$ as $\boldsymbol{p}_i$, i.e.,
\begin{align}\label{Pcolumn}
&  [\boldsymbol{p}_i ]_k= [\boldsymbol{P}_{\rm t}]_{(i-1)N_{\rm c}+k,i}, \notag\\
& \ \ \ \   k=1,2,\ldots,N_{\rm c}, ~i=1,2,\ldots,N_{\rm RF}.
\end{align}
From \eqref{Pcolumn}, if $\boldsymbol{p}_i$ is given for $i=1,2,\ldots,N_{\rm RF}$, we can determine the $i$th column of $\boldsymbol{P}_{\rm t}$ immediately.
We denote the $i$th submatrix of $\boldsymbol{S}_{\rm t}$, which is made up of the columns from $(i-1)N_{\rm c}+1$ to $iN_{\rm c}$ of $\boldsymbol{S}_{\rm t}$ as $\boldsymbol{Q}_i$, i.e.,
\begin{align}\label{Qmatrix}
& [\boldsymbol{Q}_i]_{k,l}= [\boldsymbol{S}_{\rm t}]_{k,(i-1)N_{\rm c}+l}, \notag\\
& \ \ \ \ \ \   k=1,2,\ldots,N_{\rm t}, ~l=1,2,\ldots,N_{\rm c}.
\end{align}
Based on \eqref{Qmatrix}, if $\boldsymbol{Q}_i$ for $i=1,2,\ldots,N_{\rm RF}$ is given, we can determine $\boldsymbol{S}_{\rm t}$ by
\begin{equation}\label{StDetermination}
  \boldsymbol{S}_{\rm t} = [\boldsymbol{Q}_1,\boldsymbol{Q}_2,\ldots,\boldsymbol{Q}_{N_{\rm RF}}].
\end{equation}
Then \eqref{Secondstep} can be converted into $N_{\rm RF}$ independent subproblems, where the $i$th subproblem for $i=1,2,\ldots,N_{\rm RF}$ is written as
\vspace{-0.2cm}
\begin{subequations}\label{Subproblem}
\begin{align}
& \min_{\boldsymbol{Q}_i, \boldsymbol{p}_i} {\big\|[ \widehat{\boldsymbol{F}}_{\rm RF}]_{:,i}-\boldsymbol{Q}_i\boldsymbol{p}_i\big\|}^2_F\\
&~~{\rm s.t.}  \ \  [\boldsymbol{Q}_i]_{m,n}\in \mathcal{A},~\forall m,n,\label{constraint21}\\
& \ \ \ \ \ \ \ \  \angle\big( [\boldsymbol{p}_{i}]_{k} \big)\in \mathcal{B},~\Big|[\boldsymbol{p}_i]_{k}\Big|=\frac{1}{\sqrt{N_{\rm c}}},~k=1,2,\ldots,N_{\rm c}.\label{constraint22}
\end{align}
\end{subequations}

From \eqref{Subproblem}, $\boldsymbol{Q}_i$ and $\boldsymbol{p}_i$ are coupled. Again we resort to alternating minimization to solve it.

\subsubsection{Determination of \texorpdfstring{$\boldsymbol{p}_i$}{} given \texorpdfstring{$\boldsymbol{Q}_i$}{}}
Given $\boldsymbol{Q}_i$, the determination of $\boldsymbol{p}_i$ based on \eqref{Subproblem} can be rewritten as
\begin{align}\label{Solvepi}
& \min_{\boldsymbol{p}_i} \ {\big\|[ \widehat{\boldsymbol{F}}_{\rm RF}]_{:,i}-\boldsymbol{Q}_i\boldsymbol{p}_i\big\|}^2_F \notag\\
&~ {\rm s.t.}  \ \ \angle\big( [\boldsymbol{p}_{i}]_{k} \big)\in \mathcal{B},~\Big|[\boldsymbol{p}_i]_{k}\Big|=\frac{1}{\sqrt{N_{\rm c}}},~k=1,2,\ldots,N_{\rm c}.
\end{align}

Each entries in $\boldsymbol{p}_i$ is selected from a candidate set. For the exhaustive search, it needs to find a best combination from $2^{bN_{\rm c}}$ ones. For example, if $b=3$ and $N_{\rm c}=8$, it needs to exhaustively search from $2^{24}=16777216$ combinations, implying that it is computationally inefficient. To reduce the computational complexity, we may temporarily relax the constraint in \eqref{Solvepi} by converting the discrete entries of $\boldsymbol{p}_i$ into continuous ones, which can be expressed as
\begin{align}\label{Solvepi2}
& \min_{\boldsymbol{p}_i} \ {\big\|[ \widehat{\boldsymbol{F}}_{\rm RF}]_{:,i}-\boldsymbol{Q}_i\boldsymbol{p}_i\big\|}^2_F \notag\\
&~ {\rm s.t.}  \ \ \Big|[\boldsymbol{p}_i]_k\Big|=\frac{1}{\sqrt{N_{\rm c}}},k=1,2,\ldots,N_{\rm c}.
\end{align}
Note that \eqref{Solvepi2} is a typical Riemannian manifold optimization problem and can be solved by the existing toolbox. Suppose we obtain a solution $\boldsymbol{\widetilde{p} }_i$ from \eqref{Solvepi2}. Then the phases of $\boldsymbol{\widetilde{p} }_i$ is denoted as ${\widetilde{\theta}}_{ki} \triangleq \angle\big( [\boldsymbol{\widetilde{p}}_{i}]_{k} \big)$ for $k=1,2,\ldots,N_{\rm c}$. We quantize ${\widetilde{\theta}}_{ki}$ by solving
\begin{equation}\label{Quantization}
{\widetilde{\theta}}_{ki} = \arg \min_{\theta \in \mathcal{B}} {\Big|{\widetilde{\theta}}_{ki}-\theta\Big |}
\end{equation}
so that we can obtain a feasible solution for \eqref{Solvepi}.

\begin{algorithm}[!t]
	\caption{VPS-HPD Scheme}
	\label{alg1}
	\begin{algorithmic}[1]
		\STATE \textbf{Input:} $\boldsymbol{F}_{\rm opt}$ in \eqref{SVD-Fopt}.
        \STATE Initialize $\boldsymbol{F}_{\rm BB}$ as a random full-column-rank matrix.
        \REPEAT
        \STATE Obtain $\widehat{\boldsymbol{F}}_{\rm RF}$ via (\ref{Analogm}).
        \FOR{$i=1:N_{\rm RF}$}
        \STATE Initialize $\boldsymbol{Q}_{i}$ as a random binary matrix.
        \REPEAT
        \STATE Obtain $\boldsymbol{p}_i$ via (\ref{Solvepi2}).
        \STATE Obtain $\boldsymbol{Q}_i$ via (\ref{Solveonerow}).
        \UNTIL \textit{stop condition~(1)} is satisfied
        \ENDFOR
        \STATE Obtain $\boldsymbol{P}_{\rm t}$  and $\boldsymbol{S}_{\rm t}$ via \eqref{Pcolumn} and \eqref{StDetermination}, respectively.
        \STATE Quantize $\boldsymbol{P}_{\rm t}$ via \eqref{QuantizationForPt1}.
        \STATE Obtain $\boldsymbol{F}_{\rm BB}$ by (\ref{Digitalm}).
        \UNTIL \textit{stop condition~(2)} is satisfied
        \STATE Normalize $\boldsymbol{F}_{\rm BB}$ by (\ref{Normalize}).
        \STATE \textbf{Output:} $\boldsymbol{F}_{\rm BB}$, $\boldsymbol{S}_{\rm t}$, $\boldsymbol{P}_{\rm t}$.
	\end{algorithmic}
\end{algorithm}

\subsubsection{Determination of \texorpdfstring{$\boldsymbol{Q}_i$}{} given \texorpdfstring{$\boldsymbol{p}_i$}{}}

For a given $\boldsymbol{p}_i$, (\ref{Subproblem}) can be rewritten as
\begin{align}\label{SolveAi}
& \ \min_{\boldsymbol{Q}_i} \ {\big\|[ \widehat{\boldsymbol{F}}_{\rm RF}]_{:,i}-\boldsymbol{Q}_i\boldsymbol{p}_i\big\|}^2_F \notag\\
&~~{\rm s.t.}  \ \ \  [\boldsymbol{Q}_i]_{m,n}\in \mathcal{A},~\forall m,n.
\end{align}
In fact, \eqref{SolveAi} can be converted into $N_{\rm t}$ independent subproblems, where the $m$th subproblem for $m=1,2,\ldots,N_{\rm t}$ can be expressed as
\begin{align}\label{Solveonerow}
&\min_{\ \ \ [\boldsymbol{Q}_i]_{m,:}}{\Big|[ \widehat{\boldsymbol{F}}_{\rm RF}]_{m,i}-[\boldsymbol{Q}_i]_{m,:}\boldsymbol{p}_i\Big |} \notag\\
&~~ \ \ {\rm s.t.}  \ \ \ \ [\boldsymbol{Q}_i]_{m,n}\in \mathcal{A},n=1,2,\ldots,N_{\rm c}.
\end{align}
Each subproblem can be solved by the exhaustive search to find a best combination from $2^{N_{\rm c}}$ ones. If $N_{\rm c}=8$, we needs to search from $2^8=256$ combinations, which is computationally tractable.

Given a random and binary initialization of $\boldsymbol{Q}_i$, we can iteratively run the procedures described by the equations from \eqref{Solvepi} to \eqref{Solveonerow} until \textit{stop condition~(1)} is satisfied, where \textit{stop condition~(1)}  can be set as equaling a predefined number of iterations. Then optimized $\boldsymbol{Q}_i$ and $\boldsymbol{p}_i$ can be obtained. Consequently, we can obtain $\boldsymbol{S}_{\rm t}$ and $\boldsymbol{P}_{\rm t}$ via \eqref{Qmatrix} and \eqref{Pcolumn}, respectively.

For each iteration of alternating optimization on $\boldsymbol{p}_i$ and $\boldsymbol{Q}_i$, we need to quantize each entry of $\boldsymbol{p}_i$ to satisfy the hardware constraint coming from the limited resolution of phase shifters by \eqref{Quantization}. In fact, we may move the above quantization step outside of the alternating optimization. During the alternating optimization on $\boldsymbol{p}_i$ and $\boldsymbol{Q}_i$, we use continuous phase for $\boldsymbol{p}_i$, which can remove the computational complexity of frequent quantization during the iterations. After $N_{\rm RF}$ times of alternating optimization is all finished and $\boldsymbol{P}_{\rm t}$ is obtained according to \eqref{Pcolumn}, we make quantization for each entry of $\boldsymbol{P}_{\rm t}$. Suppose the phase of $[\boldsymbol{P}_{\rm t}]_{k,i}$ is denoted as ${\widehat{\theta}}_{ki} \triangleq \angle\big( [\boldsymbol{P}_{\rm t}]_{k,i}\big)$ for $k=1,2,\ldots,N_{\rm c}N_{\rm RF}$ and $i=1,2,\ldots,N_{\rm RF}$. We quantize and update ${\widehat{\theta}}_{ki}$ by
\begin{equation}\label{QuantizationForPt1}
{\widehat{\theta}}_{ki} \leftarrow  \arg \min_{\theta \in \mathcal{B}} {\Big|{\widehat{\theta}}_{ki}-\theta\Big |}
\end{equation}
so that we can satisfy the hardware constraint coming from the limited resolution of phase shifters.

\subsection{Digital Precoder Design}\label{subsec.Matrixsolution}
Given $\boldsymbol{S}_{\rm t}$ and $\boldsymbol{P}_{\rm t}$, which are equivalent as given $\boldsymbol{F}_{\rm RF}$ according to~\eqref{newanalog}, we can compute $\boldsymbol{F}_{\rm BB}$ in \eqref{solveto2} by the least square method as
\begin{equation}\label{Digitalm}
\boldsymbol{F}_{\rm BB}=(\boldsymbol{P}_{\rm t}^H \boldsymbol{S}_{\rm t}^H\boldsymbol{S}_{\rm t}\boldsymbol{P}_{\rm t})^{-1}\boldsymbol{P}_{\rm t}^H  \boldsymbol{S}_{\rm t}^H\boldsymbol{F}_{\rm opt}.
\end{equation}

Given an random initialization of $\boldsymbol{F}_{\rm BB}$ which is full column rank, we can determine $\boldsymbol{S}_{\rm t}$ and $\boldsymbol{P}_{\rm t}$, based on which we can further obtain an optimized $\boldsymbol{F}_{\rm BB}$. In this way, we can iteratively optimize $\boldsymbol{S}_{\rm t}$, $\boldsymbol{P}_{\rm t}$ and $\boldsymbol{F}_{\rm BB}$ until \textit{stop condition (2)} is satisfied where \textit{stop condition (2)} can be set as equaling a predefined number of iterations.

Finally, to satisfy the constraint of \eqref{constraint3}, we normalize the digital precoder $\boldsymbol{F}_{\rm BB}$ as the new one by
\begin{equation}\label{Normalize}
\boldsymbol{F}_{\rm BB} \leftarrow \frac{\sqrt{N_{\rm s}}}{{\|\boldsymbol{S}_{\rm t}\boldsymbol{P}_{\rm t}\boldsymbol{F}_{\rm BB}\|}^2_F}\boldsymbol{F}_{\rm BB}.
\end{equation}

The detailed steps of the proposed VPS-HPD scheme is summarized in~\textbf{Algorithm~1}. Note that the hybrid combining design including $\boldsymbol{W}_{\rm RF}$ and $\boldsymbol{W}_{\rm BB}$ are similar.

\subsection{Convergence and Computational Complexity Analysis}\label{subsec.Comparation}
\subsubsection{Convergence}
For the VPS-HPD scheme, when using alternating optimization to solve $N_{\rm RF}$ subproblems in \eqref{Subproblem}, we adopt the Riemannian manifold optimization and exhaustive search to determine $\boldsymbol p_i$ and $\boldsymbol Q_i$, respectively. Note that the Riemannian manifold optimization can guarantee the monotonic decreasing of the objective function of \eqref{Solvepi2} in each iteration. The exhaustive search can also guarantee the result of $\boldsymbol Q_i$ is optimal in each iteration. Therefore, the convergence of the VPS-HPD scheme can be verified.

\subsubsection{Computational Complexity} 
Suppose the predefined numbers of iterations for \textit{stop condition (1)} and \textit{stop condition~(2)} are $N_{\rm max}^{(1)}$ and $N_{\rm max}^{(2)}$, The alternating optimization of the analog precoder and the digital precoder is performed for $N_{\rm max}^{(2)}$ iterations. During each of the $N_{\rm max}^{(2)}$ iterations, the computational complexity to design $\widehat{\boldsymbol{F}}_{\rm RF}$ in \eqref{Analogm} and $\boldsymbol{F}_{\rm BB}$ in \eqref{Digitalm} is $\mathcal{O}(N_{\rm RF}^3)$ and $\mathcal{O}(N_{\rm t}N_{\rm RF}^2)$, respectively. The analog precoder design is converted into $N_{\rm RF}$ subproblems, where each subproblem further includes the alternating optimization of the phase matrix and switch matrix for $N_{\rm max}^{(1)}$ iterations. When optimizing the switch matrix, it is further converted into $N_{\rm t}$ subproblems and each subproblem needs $2^{N_{\rm c}}$ iterations for the exhaustive search. Then the computational complexity for the VPS-HPD is
\begin{equation}\label{VPS-HPD-Complexity}
  \mathcal{O}\bigg(N_{\rm max}^{(2)}\big(N_{\rm RF}^4 N_{\rm max}^{(1)}(N_{\rm t}2^{N_{\rm c}} + \xi ) + N_{\rm t}N_{\rm RF}^2\big)\bigg)
\end{equation}
where $\xi$ represents the number of complex-valued multiplication to obtain a solution using the Riemannian manifold optimization. In fact, during each of the $N_{\rm max}^{(1)}$ iterations, the update of $\widetilde{\boldsymbol{p}}_i$ involves a line search algorithm, which causes the nested loops within the Riemannian manifold optimization to slow down the whole optimization procedure. Furthermore, the Kronecker products within the Riemannian manifold optimization will result in an exponential increase of the computational complexity~\cite{Xianghao2016alter}.

\section{Low-Complexity Hybrid Precoder Design}\label{sec.lowcomplexity}
The VPS-HPD scheme uses Riemannian manifold optimization to optimize $\boldsymbol{p}_i$ in \eqref{Solvepi2} and the exhaustive search to find an optimal $\boldsymbol{Q}_i$ in \eqref{Solveonerow}, which incurs computational complexity. In the following, we will address this issue and propose a new scheme without the Riemannian manifold optimization and exhaustive search.

It has been shown that imposing a semi-orthogonal structure for $\boldsymbol{F}_{\rm BB}$ is an efficient way to get near-optimal performance~\cite{yu2019hard,YuWei2016JSTSP}. The semi-orthogonal structure for $\boldsymbol{F}_{\rm BB}$ can be defined as
\begin{equation}\label{SemiOrthogonalStructure}
  \boldsymbol{F}_{\rm BB} \triangleq \alpha \boldsymbol{F}_{\rm DD}
\end{equation}
where $\alpha$ denotes a nonzero scaling factor and $\boldsymbol{F}_{\rm DD}$ is a semi-unitary matrix satisfying $\boldsymbol{F}_{\rm DD}^H \boldsymbol{F}_{\rm DD} = \boldsymbol{I}_{N_{\rm s}}$. We have
\begin{equation}\label{semio}
\boldsymbol{F}_{\rm BB}^H \boldsymbol{F}_{\rm BB}=\alpha^2 \boldsymbol{I}_{N_{\rm s}},~\alpha \neq 0.
\end{equation}

According to \textit{Lemma~1} in~\cite{yu2019hard}, the objective function of \eqref{solveto2}, denoted as ${\big\|\boldsymbol{F}_{\rm opt}-\boldsymbol{S}_{\rm t}\boldsymbol{P}_{\rm t}\boldsymbol{F}_{\rm BB}\big\|}^2_F$, is upper bounded by
\begin{equation}\label{upper}
{\big\|\boldsymbol{F}_{\rm opt}\big\|}^2_F-2\alpha {\rm{Re}}\Big( {\rm{tr}} \big( \boldsymbol{F}_{\rm DD}\boldsymbol{F}_{\rm opt}^H \boldsymbol{S}_{\rm t} \boldsymbol{P}_{\rm t}\big) \Big)+ {\alpha}^2 {\big\|\boldsymbol{S}_{\rm t}\big\|}^2_F.
\end{equation}
Adopting this upper bound as the surrogate objective, we can convert \eqref{solveto2} into
\begin{subequations}\label{newpro}
\begin{align}
& \min_{\boldsymbol{S}_{\rm t},\boldsymbol{P}_{\rm t},\alpha,\boldsymbol{F}_{\rm DD}} {\alpha}^2 {\big\|\boldsymbol{S}_{\rm t}\big\|}^2_F-2\alpha {\rm{Re}}\Big( {\rm{tr}} \left( \boldsymbol{F}_{\rm DD}\boldsymbol{F}_{\rm opt}^H \boldsymbol{S}_{\rm t} \boldsymbol{P}_{\rm t}\right)\Big)\\
&~~~~~~{\rm s.t.}  \ \ \ \ \angle\big( [\boldsymbol{P}_{\rm t}]_{i,l} \big)\in \mathcal{B},~\Big|[\boldsymbol{P}_{\rm t}]_{i,l}\Big|=\frac{1}{\sqrt{N_{\rm c}}},~\forall i,l, \label{constraintup1}\\
& \ \ \ \ \ \ \ \ \ \ \ \ \ \ [\boldsymbol{S}_{\rm t}]_{m,n}\in\mathcal{A},~\forall m,n, \label{constraintup2}\\
& \ \ \ \ \ \ \ \ \ \ \ \ \ \ \boldsymbol{F}_{\rm DD}^H \boldsymbol{F}_{\rm DD}=\boldsymbol{I}_{N_{\rm s}},
\end{align}
\end{subequations}
where the constant term ${\big\|\boldsymbol{F}_{\rm opt}\big\|}^2_F$ is independent of $\boldsymbol{S}_{\rm t}$, $\boldsymbol{P}_{\rm t}$, $\alpha$ and $\boldsymbol{F}_{\rm DD}$ and therefore is dropped. Note that the introduce of the surrogate objective may result in a small performance gap between it and the objective function of \eqref{solveto2}; but it is the cost to reduce the computational complexity of VPS-HPD.

To solve \eqref{newpro}, we resort to the alternating optimization in three stages and propose a VPS-LC-HPD scheme, where each stage has a closed-form solution and can be efficiently implemented. In the first subsection as the first stage of VPS-LC-HPD, we design $\boldsymbol{F}_{\rm DD}$ for given $\boldsymbol{S}_{\rm t}$, $\boldsymbol{P}_{\rm t}$ and $\alpha$. In the second subsection as the second stage of VPS-LC-HPD, we design  $\boldsymbol{P}_{\rm t}$ for given $\boldsymbol{S}_{\rm t}$, $\boldsymbol{F}_{\rm DD}$ and $\alpha$. In the third subsection as the third stage of VPS-LC-HPD, we design $\boldsymbol{S}_{\rm t}$ and $\alpha$ for given $\boldsymbol{P}_{\rm t}$ and $\boldsymbol{F}_{\rm DD}$. The procedures in these three stages are iteratively performed in turn. In the last subsection, we present the initialization and stop condition for VPS-LC-HPD as well as some analysis on the computational complexity.

\subsection{Design of Semi-unitary Matrix}\label{subsec.dualnorm}
Given $\boldsymbol{S}_{\rm t}$, $\boldsymbol{P}_{\rm t}$ and $\alpha$, the design of $\boldsymbol{F}_{\rm DD}$ according to \eqref{newpro} can be  expressed as
\begin{subequations}\label{solvefdd}
  \begin{align}
& \max_{\boldsymbol{F}_{\rm DD}} \ 2\alpha {\rm{Re}}\Big( {\rm{tr}} \big( \boldsymbol{F}_{\rm DD}\boldsymbol{F}_{\rm opt}^H \boldsymbol{S}_{\rm t} \boldsymbol{P}_{\rm t}\big) \Big)\\
&~~~{\rm s.t.}  \ \ \ \ \boldsymbol{F}_{\rm DD}^H \boldsymbol{F}_{\rm DD}=\boldsymbol{I}_{N_s}.
\end{align}
\end{subequations}
For the objective function of \eqref{solvefdd}, we have
\begin{equation}
\begin{aligned}\label{dualnorm}
\alpha & {\rm{Re}}\Big( {\rm{tr}} \big( \boldsymbol{F}_{\rm DD}\boldsymbol{F}_{\rm opt}^H \boldsymbol{S}_{\rm t} \boldsymbol{P}_{\rm t}\big) \Big) \leq \big|{\rm{tr}}\left(\alpha\boldsymbol{F}_{\rm DD}\boldsymbol{F}_{\rm opt}^H\boldsymbol{S}_{\rm t}\boldsymbol{P}_{\rm t}\right)\big|\\
&\stackrel{(a)}\leq \big\|\boldsymbol{F}_{\rm DD}^H\big\|_{\infty} \big\|\alpha\boldsymbol{F}_{\rm opt}^H\boldsymbol{S}_{\rm t}\boldsymbol{P}_{\rm t}\big\|_1=\big\|\alpha\boldsymbol{F}_{\rm opt}^H\boldsymbol{S}_{\rm t}\boldsymbol{P}_{\rm t}\big\|_1 \ =\sum\limits_{i=1}^{N_{\rm s}}\sigma_i.
\end{aligned}
\end{equation}
where $\|\cdot\|_{\infty}$ and $\|\cdot\|_1$ denote the infinite and one Schatten norm~\cite{MatrixAnalysis2012}, respectively; $(a)$ follows the H${\rm \ddot o}$lder's inequality; and $\sigma_i$ denotes the $i$th singular value of $\alpha\boldsymbol{F}_{\rm opt}^H\boldsymbol{S}_{\rm t}\boldsymbol{P}_{\rm t}$ for $i=1,2,\ldots,N_{\rm s}$. Denoting the truncated SVD of $\alpha\boldsymbol{F}_{\rm opt}^H\boldsymbol{S}_{\rm t}\boldsymbol{P}_{\rm t}$ as
\begin{equation}\label{SVD2}
  \alpha\boldsymbol{F}_{\rm opt}^H\boldsymbol{S}_{\rm t}\boldsymbol{P}_{\rm t} = \boldsymbol{\Phi \Theta \Omega} ^{H},
\end{equation}
where $\boldsymbol{\Phi}\in \mathbb{C}^{N_{\rm s}\times N_{\rm s}}$ is a unitary matrix and $\boldsymbol{\Omega}\in \mathbb{C}^{N_{\rm RF}\times N_{\rm s}}$ is a truncated unitary matrix. We have $\sigma_i = [\boldsymbol{\Theta}]_{i,i},i=1,2,\ldots,N_{\rm s}$. When $(a)$  holds, it requires
\begin{equation}\label{fddans}
\boldsymbol{F}_{\rm DD}=\boldsymbol{\Omega} \boldsymbol{\Phi}^H
\end{equation}
to achieves the maximum value for the objective function of~\eqref{solvefdd}.

\subsection{Design of Phase Matrix}\label{subsec.matching}
Given $\boldsymbol{S}_{\rm t}$, $\boldsymbol{F}_{\rm DD}$ and $\alpha$, the design of $\boldsymbol{P}_{\rm t}$ according to \eqref{newpro} can be  expressed as
\begin{subequations}\label{solvept}
\begin{align}
& \max_{\boldsymbol{P}_{\rm t}} \ \ \ 2\alpha {\rm{Re}}\Big( {\rm{tr}} \big( \boldsymbol{F}_{\rm DD}\boldsymbol{F}_{\rm opt}^H \boldsymbol{S}_{\rm t} \boldsymbol{P}_{\rm t}\big) \Big)\\
&~~{\rm s.t.}  \ \ \ \angle\big( [\boldsymbol{P}_{\rm t}]_{i,l} \big)\in \mathcal{B},~\Big|[\boldsymbol{P}_{\rm t}]_{i,l}\Big|=\frac{1}{\sqrt{N_{\rm c}}},~\forall i,l.
\end{align}
\end{subequations}

Since ${\rm{tr}}\left(\boldsymbol{A}\boldsymbol{B}\right)={\rm{tr}}\left(\boldsymbol{B}\boldsymbol{A}\right)$ for any matrices $\boldsymbol{A}$ and $\boldsymbol{B}$ with proper dimensions, we have
\begin{equation}
   {\rm tr} \big( \boldsymbol{F}_{\rm DD}\boldsymbol{F}_{\rm opt}^H \boldsymbol{S}_{\rm t} \boldsymbol{P}_{\rm t}\big) = {\rm tr} (\boldsymbol{F}_{\rm opt}^H \boldsymbol{S}_{\rm t}\boldsymbol{P}_{\rm t}\boldsymbol{F}_{\rm DD} ).
\end{equation}
Based on the equation~\cite{MatrixAnalysis2012}
\begin{align}\label{dengshi}
{\rm{tr}}\left(\boldsymbol{A}^H\boldsymbol{B}\boldsymbol{C}\boldsymbol{D}\right)= {\rm{vec}}^H(\boldsymbol{A})\left(\boldsymbol{D}^T\otimes \boldsymbol{B}\right){\rm {vec}}(\boldsymbol{C}),
\end{align}
for any matrices $\boldsymbol{A}$, $\boldsymbol{B}$, $\boldsymbol{C}$ and $\boldsymbol{D}$, where $\boldsymbol{D}^T \otimes \boldsymbol{B}$ denotes the Kronecker product between $\boldsymbol{D}^T$ and $\boldsymbol{B}$, we have
\begin{equation}
  {\rm tr} (\boldsymbol{F}_{\rm opt}^H \boldsymbol{S}_{\rm t}\boldsymbol{P}_{\rm t}\boldsymbol{F}_{\rm DD} ) = {\rm vec}^H(\boldsymbol{F}_{\rm opt})\left(\boldsymbol{F}_{\rm DD}^T \otimes \boldsymbol{S}_{\rm t}\right){\rm {vec}}(\boldsymbol{P}_{\rm t}).
\end{equation}

Then \eqref{solvept} can be rewritten as
\begin{subequations}\label{solvept2}
\begin{align}
& \max_{\boldsymbol{P}_{\rm t}} \ \ \ 2\alpha {\rm{Re}}\Big({\rm vec}^H(\boldsymbol{F}_{\rm opt})\left(\boldsymbol{F}_{\rm DD}^T \otimes \boldsymbol{S}_{\rm t}\right){\rm {vec}}(\boldsymbol{P}_{\rm t})\Big)\\
&~~{\rm s.t.}  \ \ \ \angle\big( [\boldsymbol{P}_{\rm t}]_{i,l} \big)\in \mathcal{B},~\Big|[\boldsymbol{P}_{\rm t}]_{i,l}\Big|=\frac{1}{\sqrt{N_{\rm c}}},~\forall i,l.
\end{align}
\end{subequations}

For the simplicity of the notation, we denote
\begin{equation}\label{Computingf}
  \boldsymbol{f}^H \triangleq {\rm{vec}}^H(\boldsymbol{F}_{\rm opt})\left(\boldsymbol{F}_{\rm DD}^T \otimes \boldsymbol{S}_{\rm t}\right)
\end{equation}
which is a constant independent of $\boldsymbol{P}_{\rm t}$. It has been shown in \eqref{newanalog} that $\boldsymbol{P}_{\rm t}$ is a generalized block diagonal matrix, where each column of $\boldsymbol{P}_{\rm t}$ has only $N_{\rm c}$ nonzero entries representing the values of $N_{\rm c}$ phase shifters. Therefore, we can drop the $(N_{\rm RF}-1)N_{\rm RF}N_{\rm c}$ zero entries in ${\rm vec}(\boldsymbol{P}_{\rm t})$ and retain its nonzero entries to form a column vector, denoted as $\boldsymbol{x}$, with the length of $N_{\rm c}N_{\rm RF}$. Accordingly, we drop the $(N_{\rm RF}-1)N_{\rm RF}N_{\rm c}$ entries of $\boldsymbol{f}^H$ corresponding to the $(N_{\rm RF}-1)N_{\rm RF}N_{\rm c}$ zero entries of ${\rm vec}(\boldsymbol{P}_{\rm t})$ and get a new vector $\tilde{ \boldsymbol{f} }^H$, with the length of $N_{\rm c}N_{\rm RF}$. Then \eqref{solvept2} can be recast as
\begin{subequations}\label{solvept3}
\begin{align}
  & \max_{\boldsymbol{x}} \ \ \ 2\alpha {\rm{Re}} (\tilde{ \boldsymbol{f} }^H\boldsymbol{x})\\
&~~{\rm s.t.}  \ \ \ \big|[\boldsymbol{x}]_i \big|=1, \angle \big( [\boldsymbol{x}]_i \big) \in \mathcal{B},~\forall i.
\end{align}
\end{subequations}

The solution to \eqref{solvept3} is
\begin{align}\label{DesignPhaseMatrixSolution}
[\boldsymbol{x}]_i =\bigg\{
\begin{array}{rcl}
e^{j \tilde{\theta}_i},~~~~      &     & {\rm if}~ {\alpha     \geq    0}\\
e^{j (\tilde{\theta}_i+\pi)},      &     & {   {\rm else}}
\end{array}
\end{align}
where
\begin{equation}\label{QuantizationForPt}
\tilde{\theta}_i =  \arg \min_{\theta \in \mathcal{B}} \Bigg| \frac{ [ \tilde{ \boldsymbol{f} } ]_i }{ \big| [\tilde{ \boldsymbol{f} } ]_i |   }  - e^{j\theta} \Big|
\end{equation}
for $i=1,2,\ldots,N_{\rm c} N_{\rm RF}$. Once $\boldsymbol{x}$ is obtained from \eqref{DesignPhaseMatrixSolution}, $\boldsymbol{P}_{\rm t}$ is determined.

\newcounter{TempEqCnt} 
\setcounter{TempEqCnt}{\value{equation}} 
\setcounter{equation}{43} 
\begin{figure*}[t]
	\begin{small}
		\begin{align}\label{DerivationObjectiveC}
			& {\alpha}^2 {\big\|\boldsymbol{S}_{\rm t}\big\|}^2_F-2\alpha {\rm{Re}}\Big( {\rm{tr}} ( \boldsymbol{F}_{\rm DD}\boldsymbol{F}_{\rm opt}^H \boldsymbol{S}_{\rm t} \boldsymbol{P}_{\rm t} )\Big) + \big\|{\rm Re}\left(\boldsymbol{F}_{\rm opt}\boldsymbol{F}_{\rm DD}^H\boldsymbol{P}_{\rm t}^H\right)\big\|_F^2  \nonumber \\
			\stackrel{(b)}
			= & {\alpha}^2 {\rm{tr}}(\boldsymbol{S}_{\rm t}^H \boldsymbol{S}_{\rm t})-2\alpha {\rm{Re}}\Big( {\rm{tr}} ( \boldsymbol{P}_{\rm t}\boldsymbol{F}_{\rm DD}\boldsymbol{F}_{\rm opt}^H \boldsymbol{S}_{\rm t})\Big)+{\rm{tr}}\Big({\rm Re}^T\left(\boldsymbol{F}_{\rm opt}\boldsymbol{F}_{\rm DD}^H\boldsymbol{P}_{\rm t}^H\right){\rm Re}\left(\boldsymbol{F}_{\rm opt}\boldsymbol{F}_{\rm DD}^H\boldsymbol{P}_{\rm t}^H\right)\Big)  \nonumber \\
			= & {\alpha}^2 {\rm{tr}}(\boldsymbol{S}_{\rm t}^H \boldsymbol{S}_{\rm t})-\alpha {\rm{Re}}\Big( {\rm tr} \big( (\boldsymbol{F}_{\rm opt}\boldsymbol{F}_{\rm DD}^H\boldsymbol{P}_{\rm t}^H)^H\boldsymbol{S}_{\rm t}\big)\Big)-\alpha {\rm{Re}}\Big( {\rm{tr}} ( \boldsymbol{S}_{\rm t}^H \boldsymbol{F}_{\rm opt}\boldsymbol{F}_{\rm DD}^H\boldsymbol{P}_{\rm t}^H )\Big)+{\rm{tr}}\Big({\rm Re}^T\left(\boldsymbol{F}_{\rm opt}\boldsymbol{F}_{\rm DD}^H\boldsymbol{P}_{\rm t}^H\right) {\rm Re}\left(\boldsymbol{F}_{\rm opt}\boldsymbol{F}_{\rm DD}^H\boldsymbol{P}_{\rm t}^H\right)\Big)  \nonumber \\
			= & {\alpha}^2 {\rm{tr}}(\boldsymbol{S}_{\rm t}^H \boldsymbol{S}_{\rm t})-\alpha {\rm{tr}}\Big({\rm{Re}}( (\boldsymbol{F}_{\rm opt}\boldsymbol{F}_{\rm DD}^H\boldsymbol{P}_{\rm t}^H)^H)\boldsymbol{S}_{\rm t}\Big)-\alpha {\rm{tr}}\Big(\boldsymbol{S}_{\rm t}^H {\rm{Re}}(\boldsymbol{F}_{\rm opt}\boldsymbol{F}_{\rm DD}^H\boldsymbol{P}_{\rm t}^H )\Big)+{\rm{tr}}\Big({\rm Re}^T\left(\boldsymbol{F}_{\rm opt}\boldsymbol{F}_{\rm DD}^H\boldsymbol{P}_{\rm t}^H\right) {\rm Re}\left(\boldsymbol{F}_{\rm opt}\boldsymbol{F}_{\rm DD}^H\boldsymbol{P}_{\rm t}^H\right)\Big) \nonumber \\
			= & {\rm{tr}}\bigg(\Big( {\rm Re}\big(\boldsymbol{F}_{\rm opt}\boldsymbol{F}_{\rm DD}^H\boldsymbol{P}_{\rm t}^H\big)-\alpha \boldsymbol{S}_{\rm t}\Big)^H \Big( {\rm Re}\big(\boldsymbol{F}_{\rm opt}\boldsymbol{F}_{\rm DD}^H\boldsymbol{P}_{\rm t}^H\big)-\alpha \boldsymbol{S}_{\rm t}\Big)\bigg) = \Big\|{\rm{Re}} \left(\boldsymbol{F}_{\rm opt}\boldsymbol{F}_{\rm DD}^H\boldsymbol{P}_{\rm t}^H\right)-\alpha \boldsymbol{S}_{\rm t}\Big\|_F^2.
		\end{align}
	\end{small}
	\hrulefill
\end{figure*}
\setcounter{equation}{\value{TempEqCnt}} 

\subsection{Design of Switch Matrix and Scaling Factor}\label{subsec.sw}
Given $\boldsymbol{P}_{\rm t}$ and $\boldsymbol{F}_{\rm DD}$, the design of $\boldsymbol{S}_{\rm t}$ and $\alpha$ can be addressed as follows.

By adding a constant term $\big\|{\rm Re}\left(\boldsymbol{F}_{\rm opt}\boldsymbol{F}_{\rm DD}^H\boldsymbol{P}_{\rm t}^H\right)\big\|_F^2$ that is independent of $\boldsymbol{S}_{\rm t}$ and $\alpha$ to the objective in \eqref{newpro}, we can derive \eqref{DerivationObjectiveC}, where $(b)$ follows the equations that ${\rm{tr}}\left(\boldsymbol{A}\boldsymbol{B}\right)={\rm{tr}}\left(\boldsymbol{B}\boldsymbol{A}\right)$ and $\|\boldsymbol{A}\|_F^2 = {\rm tr}(\boldsymbol{A}^H \boldsymbol{A}) $ for any matrices $\boldsymbol{A}$ and $\boldsymbol{B}$.

Based on \eqref{DerivationObjectiveC}, the optimization problem \eqref{newpro} can be rewritten as
\setcounter{equation}{44}
\begin{subequations}\label{solves1}
\begin{align}
& \min_{\boldsymbol{S}_{\rm t},\alpha} \ \ \big\|{\rm{Re}} \left(\boldsymbol{F}_{\rm opt}\boldsymbol{F}_{\rm DD}^H\boldsymbol{P}_{\rm t}^H\right)-\alpha \boldsymbol{S}_{\rm t}\big\|_F^2 \label{ObjSolves1}\\
&~~{\rm s.t.}  \ \ \ [\boldsymbol{S}_{\rm t}]_{m,n}\in\mathcal{A},~\forall m,n.
\end{align}
\end{subequations}

It is essentially an entry-wise minimization problem, where we aim to minimize the absolute value of the difference between each entry of ${\rm{Re}} \left(\boldsymbol{F}_{\rm opt}\boldsymbol{F}_{\rm DD}^H\boldsymbol{P}_{\rm t}^H\right)$ and the corresponding entry of $\alpha \boldsymbol{S}_{\rm t}$. If $\alpha >0$, we may determine each entry of $ \boldsymbol{S}_{\rm t}$ by
\begin{align}\label{DesignPhaseMatrixSolution2}
[\boldsymbol{S}_{\rm t}]_{m,n} =\bigg\{
\begin{array}{rcl}
1,      &      {\rm if}~ { \big[{\rm{Re}} (\boldsymbol{F}_{\rm opt}\boldsymbol{F}_{\rm DD}^H\boldsymbol{P}_{\rm t}^H )\big]_{m,n}     \geq    \alpha/2}\\
0,      &      {   {\rm else}}~~~~~~~~~~~~~~~~~~~~~~~~~~~~~~~~~~~
\end{array}
\end{align}
for $m=1,2,\dots, N_{\rm t}$ and $n=1,2,\ldots,N_{\rm c} N_{\rm RF}$. If $\alpha <0$, we may determine each entry of $ \boldsymbol{S}_{\rm t}$ by
\begin{align}\label{DesignPhaseMatrixSolution3}
[\boldsymbol{S}_{\rm t}]_{m,n} =\bigg\{
\begin{array}{rcl}
0,      &      {\rm if}~ { \big[{\rm{Re}} (\boldsymbol{F}_{\rm opt}\boldsymbol{F}_{\rm DD}^H\boldsymbol{P}_{\rm t}^H )\big]_{m,n}     \geq    \alpha/2}\\
1,      &      {   {\rm else}}~~~~~~~~~~~~~~~~~~~~~~~~~~~~~~~~~~~
\end{array}
\end{align}
for $m=1,2,\dots, N_{\rm t}$ and $n=1,2,\ldots,N_{\rm c} N_{\rm RF}$. Therefore, $\boldsymbol{S}_{\rm t}$ can be determined if $\alpha$ is given. In the following, we will elaborate on how to determine $\alpha$.

By vectorizing the matrices in \eqref{ObjSolves1}, we convert \eqref{solves1} to
\begin{subequations}\label{solves12}
\begin{align}
& \min_{\boldsymbol{S}_{\rm t},\alpha} \ \ \big\| {\rm vec} \big( {\rm{Re}} (\boldsymbol{F}_{\rm opt}\boldsymbol{F}_{\rm DD}^H\boldsymbol{P}_{\rm t}^H )\big)-\alpha {\rm vec} (\boldsymbol{S}_{\rm t})\big\|_2^2\\
&~~{\rm s.t.}  \ \ \ [\boldsymbol{S}_{\rm t}]_{m,n}\in\mathcal{A},~\forall m,n.
\end{align}
\end{subequations}

We sort the real-valued vector ${\rm vec} \big( {\rm{Re}} (\boldsymbol{F}_{\rm opt}\boldsymbol{F}_{\rm DD}^H\boldsymbol{P}_{\rm t}^H )\big)$ in the ascending order, obtaining a new column vector $\boldsymbol{z}\triangleq[\boldsymbol{z}_1, \boldsymbol{z}_2, \ldots,\boldsymbol{z}_N]^T$ with the length $N \triangleq N_{\rm t}N_{\rm c}N_{\rm RF}$. The entries of $\boldsymbol{z}$ satisfy $\boldsymbol{z}_1  \leq \boldsymbol{z}_2 \leq \ldots \leq \boldsymbol{z}_N$, which divides the real-valued axis into $N+1$ intervals, including $(-\infty,\boldsymbol{z}_1]$, $(\boldsymbol{z}_1,\boldsymbol{z}_2]$, $(\boldsymbol{z}_2,\boldsymbol{z}_3]$, \ldots, $(\boldsymbol{z}_{N-1},\boldsymbol{z}_N]$, $(\boldsymbol{z}_{N},+\infty)$. We will discuss $\alpha/2$ lies in each of the $N+1$ intervals case by case, since different value of $\alpha/2$ leads to different $\boldsymbol{S}_{\rm t}$ according to \eqref{DesignPhaseMatrixSolution2} and \eqref{DesignPhaseMatrixSolution3}. If the entries of $\boldsymbol{z}$ happens to be the same, we just skip the corresponding intervals. Note that if $\alpha/2$ lies in $(-\infty,\boldsymbol{z}_1]$ or $(\boldsymbol{z}_{N},+\infty)$, the entries of $\boldsymbol{S}_{\rm t}$ will be all zeros or all ones, which indicates that all signals are switched off or switched on, respectively, and is therefore impractical. If $\alpha/2$ lies in the interval $\mathcal{L}_i \triangleq (\boldsymbol{z}_i,\boldsymbol{z}_{i+1}]$ equivalently having $2\boldsymbol{z}_i < \alpha \leq 2\boldsymbol{z}_{i+1}$ for $i=1,2,\ldots,N-1$, \eqref{ObjSolves1} can be defined as
\begin{align}\label{solvept9}
&g(\alpha,i) \triangleq \big\| {\rm vec} \big( {\rm{Re}} (\boldsymbol{F}_{\rm opt}\boldsymbol{F}_{\rm DD}^H\boldsymbol{P}_{\rm t}^H )\big)-\alpha {\rm vec} (\boldsymbol{S}_{\rm t}) \big\|_2^2\\
&=\left\{
\begin{small}
\begin{aligned}
\sum_{m=1}^i (\boldsymbol{z}_m-\alpha)^2+\sum_{m=i+1}^{N}\boldsymbol{z}^2_m,  ~~\alpha<0,\alpha/2 \in \mathcal{L}_i,   \\
\sum_{m=1}^i \boldsymbol{z}_m^2+\sum_{m=i+1}^{N}(\boldsymbol{z}_m-\alpha)^2,  ~~\alpha>0, \alpha/2 \in \mathcal{L}_i,
\end{aligned}
\end{small}
\right.\\
&=\left\{
\begin{small}
\begin{aligned} \label{QuadraticFunction}
i\alpha^2- 2\alpha\sum_{m=1}^{i}\boldsymbol{z}_m + \sum_{m=1}^{N}\boldsymbol{z}^2_m, ~\alpha<0,\alpha/2 \in \mathcal{L}_i \\
(N-i)\alpha^2-2\alpha\sum_{m=i+1}^{N}\boldsymbol{z}_m+\sum_{m=1}^{N}\boldsymbol{z}^2_m,   ~\alpha>0,\alpha/2 \in \mathcal{L}_i
\end{aligned}
\end{small}
\right.
\end{align}

For two quadratic functions in terms of $\alpha$ in \eqref{QuadraticFunction}, the vertexes of the parabolas correspond to $\alpha=\alpha^*_i$ where
\begin{equation}
\label{solves6}
\alpha^*_i \triangleq \left\{
\begin{aligned}
\frac{\sum\nolimits_{m=1}^{i}\boldsymbol{z}_m}{i}, ~~~~ & \alpha<0,\alpha/2 \in \mathcal{L}_i \\
\frac{\sum\nolimits_{m=i+1}^{N}\boldsymbol{z}_m}{N-i},~~  & \alpha>0,\alpha/2 \in \mathcal{L}_i .
\end{aligned}
\right.
\end{equation}

Then \eqref{solves1} can be converted to
\begin{equation}\label{solves5}
\min_{i=1,2,\ldots,N-1} \big\{ g(\alpha^*_i,i), g(2\boldsymbol{z}_i,i), g(2\boldsymbol{z}_{i+1},i)\big\}.
\end{equation}

In the following, we will show that the minimum of $g(\alpha,i)$ cannot be obtained at $\alpha=2\boldsymbol{z}_i $ or $ \alpha=2\boldsymbol{z}_{i+1}$ and can only be obtained at $\alpha=\alpha^*_i$.

If the minimum of $g(\alpha,i)$ is obtained at $\alpha=2\boldsymbol{z}_{i+1} $ instead of $\alpha=\alpha^*_i$, we have
\begin{equation}\label{Contradiction1}
  g(2\boldsymbol{z}_{i+1},i) < g(\alpha^*_k,k), \forall k=1,2,\ldots,N-1.
\end{equation}
Considering the convexity of $g(\alpha,i)$, we have $2\boldsymbol{z}_{i+1} < \alpha^*_i$. Since the entries of $\boldsymbol{z}$ are placed in ascending order, we have $\alpha^*_i \leq \alpha^*_{i+1}$ based on \eqref{solves6} and therefore $2\boldsymbol{z}_{i+1} < \alpha^*_{i+1}$. Consequently, for $g(\alpha^*_{i+1},i+1)$ obtained in the interval $\mathcal{L}_{i+1}$, we have
\begin{equation}\label{Contradiction2}
 g(\alpha^*_{i+1},i+1) <  g(2\boldsymbol{z}_{i+1},i),
\end{equation}
which contradicts with \eqref{Contradiction1}. Therefore, the minimum of $g(\alpha,i)$ can only be obtained at $\alpha=\alpha^*_i$. Then the optimal solution of \eqref{solves5} can be expressed as
\begin{equation}\label{solves6iStar}
i^* \triangleq \arg\min_{i=1,2,\ldots,N-1}  g(\alpha^*_i,i).
\end{equation}
The optimal solution of $\alpha$ is $\alpha^*_{i^*}$. By replacing $\alpha$ in \eqref{DesignPhaseMatrixSolution2} or \eqref{DesignPhaseMatrixSolution3} with $\alpha^*_{i^*}$, we can get the optimal $\boldsymbol{S}_{\rm t}$.

\subsection{Initialization and Computational Complexity}\label{subsec.ini}
The procedures in the above three stages are iteratively performed in turn until \textit{stop condition (3)} is satisfied, where \textit{stop condition (3)} can be set as equaling a predefined number of iterations. The procedures of the proposed VPS-LC-HPD scheme are summarized in \textbf{Algorithm~2}.

To initialize \textbf{Algorithm~2}, we set $\boldsymbol{S}_{\rm t}$ as a random binary matrix. Since most entries of $\boldsymbol{P}_{\rm t}$ is zero, we only need to initialize $N_{\rm c}$ nonzero entries for each column of $\boldsymbol{P}_{\rm t}$, where the phase of the $i$th nonzero entry is initialized to be $2\pi i/N_{\rm c}$ for $i=1,2,\ldots,N_{\rm c}$. We initialize $\alpha=1$, implying that we initially set $\boldsymbol{F}_{\rm BB} = \boldsymbol{F}_{\rm DD}$ according to \eqref{SemiOrthogonalStructure} for simplicity.

\begin{algorithm}[!t]
	\caption{VPS-LC-HPD Scheme}
	\label{alg2}
	\begin{algorithmic}[1]
		\STATE \textbf{Input:} $\boldsymbol{F}_{\rm opt}$ in \eqref{SVD-Fopt}.
        \STATE Initialize $\alpha$, $\boldsymbol{F}_{\rm DD}$, $\boldsymbol{S}_{\rm t}$ and $\boldsymbol{P}_{\rm t}$.
        \REPEAT
        \STATE Given $\boldsymbol{S}_{\rm t}$, $\boldsymbol{P}_{\rm t}$ and $\alpha$, we update $\boldsymbol{F}_{\rm DD}$ via (\ref{fddans}).
        \STATE Given $\boldsymbol{S}_{\rm t}$, $\boldsymbol{F}_{\rm DD}$ and $\alpha$, we update $\boldsymbol{P}_{\rm t}$ via (\ref{DesignPhaseMatrixSolution}).
        \STATE Given $\boldsymbol{P}_{\rm t}$ and $\boldsymbol{F}_{\rm DD}$, we update $\alpha$ via \eqref{solves6iStar} and \eqref{solves6}, and then update $\boldsymbol{S}_{\rm t}$ via \eqref{DesignPhaseMatrixSolution2} or \eqref{DesignPhaseMatrixSolution3}.
        \UNTIL \textit{stop condition~(3)} is satisfied
        \STATE Normalize $\boldsymbol{F}_{\rm BB}$ via (\ref{Normalize}).
        \STATE \textbf{Output:} $\boldsymbol{F}_{\rm BB}$, $\boldsymbol{S}_{\rm t}$, $\boldsymbol{P}_{\rm t}$.
	\end{algorithmic}
\end{algorithm}

\begin{figure*}[!t]
\centering
\includegraphics[width=165mm]{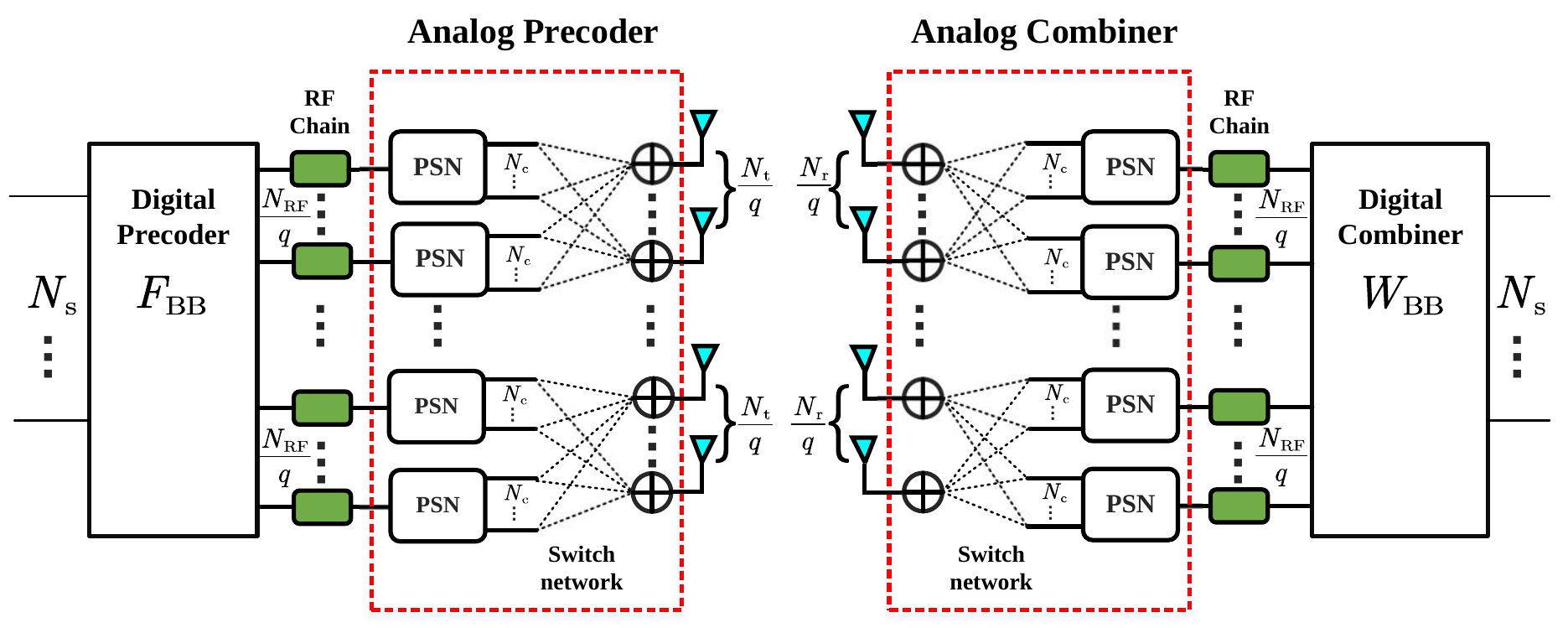}
\caption{A point-to-point mmWave massive MIMO system with the GC-VPS architecture.}
\label{fig:gs}
\end{figure*}

Now we analyze the computational complexity of the VPS-LC-HPD scheme. Suppose the predefined numbers of iterations for \textit{stop condition (3)} is $N_{\rm max}^{(3)}$. During each of $N_{\rm max}^{(3)}$ iterations, the computational complexity to update $\boldsymbol{F}_{\rm DD}$ in Subsection~\ref{subsec.dualnorm} is $\mathcal{O}(N_{\rm s}N_{\rm t}N_{\rm RF}N_{\rm c})$, which is dominated by the matrices multiplication on the left side of~\eqref{SVD2}. The computational complexity to update $\boldsymbol{P}_{\rm t}$ in Subsection~\ref{subsec.matching} is $\mathcal{O}(N_{\rm t} N_{\rm s} N_{\rm c} N_{\rm RF}^2)$, which is dominated by the matrices multiplication in~\eqref{Computingf}. The computational complexity to update $\alpha$ and $\boldsymbol{S}_{\rm t}$ in Subsection~\ref{subsec.sw} is $\mathcal{O}(N_{\rm c}N_{\rm RF}N_{\rm t}\log_2 ( N_{\rm c}N_{\rm RF}N_{\rm t}))$, which is dominated by the sorting operation for ${\rm vec} \big( {\rm{Re}} (\boldsymbol{F}_{\rm opt}\boldsymbol{F}_{\rm DD}^H\boldsymbol{P}_{\rm t}^H )\big)$. Therefore, the total computational complexity of the VPS-LC-HPD scheme is
\begin{align}\label{LC-VPS-HPDcomplexity}
  \mathcal{O}\Big(N_{\rm max}^{(3)}\big(N_{\rm s}N_{\rm t} & N_{\rm RF}N_{\rm c} + N_{\rm t} N_{\rm s} N_{\rm c} N_{\rm RF}^2 \nonumber \\
  & +N_{\rm c}N_{\rm RF}N_{\rm t}\log_2 ( N_{\rm c}N_{\rm RF}N_{\rm t})\big)\Big)
\end{align}

\section{Group-Connected VPS Architecture and Hybrid Precoding Design}\label{sec.group}
By using the low-cost switch networks, the number of phase shifters for the FPS or VPS architectures can be substantially reduced compared to the fully-connected or partially-connected architectures. To further reduce the number of switches and therefore the hardware complexity, inspired by the partially-connected architectures, we consider a GC-VPS architecture.

For the FPS or VPS architectures shown in Fig.~\ref{fig:system}, each phase shifter is connected to $N_{\rm t}$ antennas through $N_{\rm t}$ switches. However, for the GC-VPS architecture as shown in Fig.~\ref{fig:gs}, each phase shifter is only partially connected to $N_{\rm t}/q$ antennas through $N_{\rm t}/q$ switches, where we divide the $N_{\rm t}$ antennas into $q$ groups. To guarantee that each group contains at least a PSN, it is required that $q\leq N_{\rm RF}$. For simplicity in hardware implementation, we usually set that both $N_{\rm t}$ and $N_{\rm RF}$ are divisible by $q$. As a result, we can reduce the total number of switches from $N_{\rm t} N_{\rm RF} N_{\rm c}$ to $N_{\rm t} N_{\rm RF} N_{\rm c}/q$.

Since each PSN is only connected to $N_{\rm t}/q$ antennas, the analog precoder $\boldsymbol{F}_{\rm RF}$ can be divided into $q$ independent small-size precoders and can be written as a block diagonal matrix
\begin{equation}\label{blockrf}
\boldsymbol{F}_{\rm RF}=\begin{bmatrix}
\boldsymbol{F}_{{\rm RF},1} & \boldsymbol{0} &\dots & \boldsymbol{0}\\
\boldsymbol{0} & \boldsymbol{F}_{{\rm RF},2} &\dots & \boldsymbol{0}\\
\vdots &  \vdots &\ddots & \vdots \\
\boldsymbol{0} & \boldsymbol{0} & \ldots & \boldsymbol{F}_{{\rm RF},q}
\end{bmatrix},
\end{equation}
where $\boldsymbol{F}_{{\rm RF},k}\in \mathbb{C}^{\frac{N_{\rm t}}{q} \times \frac{N_{\rm RF}}{q}}$ represents the analog precoder in the $k$th group for $k=1,2,\ldots,q$. According to (\ref{newanalog}), we have
\begin{equation}\label{blockSt}
\boldsymbol{S}_{\rm t}=\begin{bmatrix}
\boldsymbol{S}_{1} & \boldsymbol{0} &\dots & \boldsymbol{0}\\
\boldsymbol{0} & \boldsymbol{S}_{2} &\dots & \boldsymbol{0}\\
\vdots &  \vdots &\ddots & \vdots \\
\boldsymbol{0} & \boldsymbol{0} & \ldots & \boldsymbol{S}_{q}
\end{bmatrix},
\end{equation}
\begin{equation}\label{blockPt}
\boldsymbol{P}_{\rm t}=\begin{bmatrix}
\boldsymbol{P}_{1} & \boldsymbol{0} &\dots & \boldsymbol{0}\\
\boldsymbol{0} & \boldsymbol{P}_{2} &\dots & \boldsymbol{0}\\
\vdots &  \vdots &\ddots & \vdots \\
\boldsymbol{0} & \boldsymbol{0} & \ldots & \boldsymbol{P}_{q}
\end{bmatrix},
\end{equation}
and
\begin{equation}\label{F_RF_Each_Group}
  \boldsymbol{F}_{{\rm RF},k}= \boldsymbol{S}_{k}\boldsymbol{P}_{k},~k=1,2,\ldots,q,
\end{equation}
where $\boldsymbol{S}_{k}\in \mathbb{Z}^{\frac{N_{\rm t}}{q} \times \frac{N_{\rm RF} N_{\rm c} }{q}}$ and $\boldsymbol{P}_{k} \in \mathbb{C}^{ \frac{N_{\rm RF} N_{\rm c} }{q} \times \frac{N_{\rm RF}}{q}}$ represent a binary switch matrix and a phase matrix in each group, respectively.

Suppose we divide $\boldsymbol{F}_{\rm BB}$ into $q$ equal-size submatrices as
\begin{equation}\label{F_BB_Submatrices}
  \boldsymbol{F}_{\rm BB} = [\boldsymbol{B}_1^T,\boldsymbol{B}_2^T,\ldots,\boldsymbol{B}_q^T]^T,
\end{equation}
where $\boldsymbol{B}_k \in \mathbb{C}^{\frac{N_{\rm RF}}{q} \times N_{\rm s}}$ represents the $k$th submatrix of $\boldsymbol{F}_{\rm BB}$ by selecting its rows from the $((k-1)N_{\rm RF} /q+1)$th row to the $(k N_{\rm RF} /q)$th row, for $k=1,2,\ldots,q$. We have
\begin{align}\label{block3}
\boldsymbol{F}_{\rm RF}\boldsymbol{F}_{\rm BB}=\big[(\boldsymbol{S}_1\boldsymbol{P}_1\boldsymbol{B}_{1})^T, (\boldsymbol{S}_2\boldsymbol{P}_2\boldsymbol{B}_{2})^T, \dots,(\boldsymbol{S}_q\boldsymbol{P}_q\boldsymbol{B}_{q})^T\big]^T.
\end{align}

We divide $\boldsymbol{F}_{\rm opt}$ into $q$ equal-size submatrices as
\begin{equation}\label{F_opt_Submatrices}
  \boldsymbol{F}_{\rm opt} = [\boldsymbol{F}_{{\rm opt},1}^T,\boldsymbol{F}_{{\rm opt},2}^T,\ldots,\boldsymbol{F}_{{\rm opt},q}^T]^T,
\end{equation}
where $\boldsymbol{F}_{{\rm opt}, k}\in \mathbb{C}^{\frac{N_{\rm t}}{q} \times N_{\rm s}}$ represents the $k$th submatrix of $\boldsymbol{F}_{\rm opt}$ by selecting its rows from the $((k-1)N_{\rm t} /q+1)$th row to the $(k N_{\rm t} /q)$th row for $k=1,2,\ldots,q$.

Then \eqref{solveto2} can be converted into $q$ independent subproblems, where the $k$th subproblem for $k=1,2,\ldots,q$ can be written as
\begin{subequations}\label{solvelast}
\begin{align}
& \min_{\boldsymbol{S}_{k},\boldsymbol{P}_{k},\boldsymbol{B}_k} {\big\|\boldsymbol{F}_{{\rm opt}, k}-\boldsymbol{S}_{k} \boldsymbol{P}_{k} \boldsymbol{B}_k\big\|}^2_F\\
&~~~~{\rm s.t.}  \ \ \ \ \angle\big( [\boldsymbol{P}_{k}]_{i,l} \big)\in \mathcal{B},~\Big|[\boldsymbol{P}_k]_{i,l}\Big|=\frac{1}{\sqrt{N_{\rm c}}},~\forall i,l, \label{constraintlast1}\\
& \ \ \ \ \ \ \ \ \ \ \ \ [\boldsymbol{S}_{k}]_{m,n}\in\mathcal{A},~\forall m,n. \label{constraintlast2}
\end{align}
\end{subequations}

By substituting $\boldsymbol{F}_{\rm opt}$ with $\boldsymbol{F}_{{\rm opt},k}$, we can run \textbf{Algorithm~1} or \textbf{Algorithm~2} to solve \eqref{solvelast}. With the obtained $\boldsymbol{S}_k$, $\boldsymbol{P}_k$ and $\boldsymbol{B}_k$ for $k=1,2,\ldots,q$, we can get $\boldsymbol{S}_{\rm t}$, $\boldsymbol{P}_{\rm t}$ and $\boldsymbol{F}_{\rm BB}$ via \eqref{blockSt},\eqref{blockPt} and \eqref{F_BB_Submatrices}, respectively.

The procedures of the proposed GC-VPS-HPD or GC-VPS-LC-HPD schemes are summarized in~\textbf{Algorithm~3}, where the former or the latter corresponds to solving \eqref{solvelast} with \textbf{Algorithm~1} or \textbf{Algorithm~2}, respectively.

\begin{algorithm}[!t]
	\caption{GC-VPS-HPD / GC-VPS-LC-HPD Scheme}
	\label{alg3}
	\begin{algorithmic}[1]
		\STATE \textbf{Input:} $\boldsymbol{F}_{\rm opt}$ in \eqref{SVD-Fopt}.
        \STATE Obtain $\boldsymbol{F}_{{\rm opt},k},k=1,2,\ldots,q$ via (\ref{F_opt_Submatrices}).
        \FOR{$k=1: q$}
        \STATE Substitute $\boldsymbol{F}_{\rm opt}$ with $\boldsymbol{F}_{{\rm opt},k}$ in \textbf{Algorithm~1} or \textbf{Algorithm~2} and then run it to obtain $\boldsymbol{S}_k$, $\boldsymbol{P}_k$ and $\boldsymbol{B}_k$.
        \ENDFOR
        \STATE Obtain $\boldsymbol{S}_{\rm t}$, $\boldsymbol{P}_{\rm t}$ and $\boldsymbol{F}_{\rm BB}$ via \eqref{blockSt}, \eqref{blockPt} and \eqref{F_BB_Submatrices}, respectively.
        \STATE Normalize $\boldsymbol{F}_{\rm BB}$ via (\ref{Normalize}).
        \STATE \textbf{Output:} $\boldsymbol{F}_{\rm BB}$, $\boldsymbol{S}_{\rm t}$, $\boldsymbol{P}_{\rm t}$.
	\end{algorithmic}
\end{algorithm}

By introducing $q$, we can balance between the system performance and hardware complexity. If $q=1$, there is only one group and each RF chain connects to all antennas, which is essentially the VPS architecture. As $q$ increases, the system performance will fall but the hardware complexity will also drop. When $q=N_{\rm RF}$, we will achieve the lowest hardware complexity.


\section{Simulation Results}\label{sec.Simulation}
The considered mmWave massive MIMO system includes a transmitter equipped with $N_{\rm t}=64$ antennas and a receiver equipped with $N_{\rm r}=16$ antennas. For both the transmitter and receiver, we use $N_{\rm RF}=4$ RF chains to support $N_{\rm s}=4$ independent data steams. The resolution of phase shifters is $b = 3$ bits, which can be set larger to meet the demand for better performance. The mmWave MIMO channel matrix is generated based on the Saleh-Valenzuela model, where the number of resolvable channel paths is set to be $L=4$, with $\alpha_{1}\sim \mathcal{CN}(0,1)$ and $\alpha_l\sim \mathcal{CN}(0,0.1)$ for $l=2,3,4$~\cite{Alk2014channelesti}.

\subsection{Spectral Efficiency Comparison}\label{subsec.simu1}
The spectral efficiency (SE) can be computed by
\begin{align}\label{SEE}
	&  \Gamma = \log_2 \big(\big |\boldsymbol{I}_{N_{\rm s}}+\frac{P}{N_{\rm s}}\boldsymbol{R}^{-1}\boldsymbol{W}_{\rm BB}^H \boldsymbol{W}_{\rm RF}^H \boldsymbol H \boldsymbol{F}_{\rm RF} \boldsymbol{F}_{\rm BB} \notag\\
	&  \ \ \ \ \ \ \ \times \boldsymbol{F}_{\rm BB}^H \boldsymbol{F}_{\rm RF}^H \boldsymbol H^H \boldsymbol{W}_{\rm RF} \boldsymbol{W}_{\rm BB} \big | \big),
\end{align}
where $\boldsymbol{R} \triangleq \sigma^2 \boldsymbol{W}_{\rm BB}^H \boldsymbol{W}_{\rm RF}^H \boldsymbol{W}_{\rm RF} \boldsymbol{W}_{\rm BB}$ is the noise covariance matrix after combining. Here we suppose that the CSI is perfectly obtained, since we want to remove the impact of the imperfect CSI on the hybrid precoding schemes for comparisons. We evaluate the SE of different HPD schemes for different architectures, including the VPS-HPD and VPS-LC-HPD schemes for the VPS architecture, the FPS-AltMin scheme for the FPS architecture~\cite{yu2019hard}, the MO-AltMin scheme for the fully-connected architecture~\cite{Xianghao2016alter}, and the fully digital precoding. To the best knowledge of the authors, the MO-AltMin scheme is the best one among all hybrid precoding schemes for the fully-connected architecture. Since the fully digital precoding needs the same number of RF chains as that of the antennas and therefore is hardware-expensive, it is only used as the performance upper bound. For fair comparisons between VPS-HPD and FPS-AltMin, we set $N_{\rm c}=8$. Moreover, we define the signal-to-noise ratio (SNR) to be $P/\sigma^2$ according to \eqref{system model}.

\begin{table*}[htb]
\centering
\caption{power consumption comparison for different mmWave MIMO architectures.}
\label{Table2}
\begin{tabular}{p{3.3cm}p{4.6cm}p{3.6cm}p{2.0cm}}
\toprule
Different architectures     &   Number / Power of phase shifters & Number / Power of switches   &  Total power \\
\midrule
Fully-connected             & $320~/~9.6\ {\rm W}$             &  $ ~~~~0~/~0 \ {\rm W}$       &  $9.6\ {\rm W}$  \\
FPS / VPS                   & $~64~/~1.92\ {\rm W}$            &  $ 2560~/~2.56\ {\rm W}$     &  $4.48\ {\rm W}$  \\
GC-VPS with $q=2$           & $~64~/~1.92\ {\rm W}$            &  $1280~/~1.28\ {\rm W}$     &  $3.2\ {\rm W}$  \\
GC-VPS with $q=4$           & $~64~/~1.92\ {\rm W}$            &  $\ 640~/~~0.64\ {\rm W}$     &  $2.56\ {\rm W}$  \\
\bottomrule
\end{tabular}
\end{table*}

\begin{figure}[!t]
\centering
\includegraphics[width=115mm]{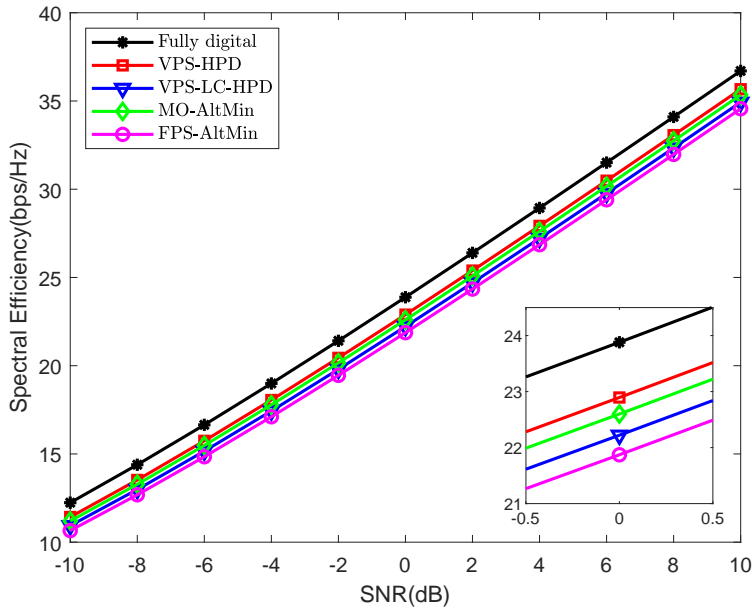}
\caption{SE comparisons of the VPS architecture with the other ones.}
\label{fig:DIS}
\end{figure}

As shown in Fig.~\ref{fig:DIS}, VPS-HPD achieves slightly better SE than that of MO-AltMin, but with much fewer phase shifters and power consumption according to Table~\ref{Table2}, owing to the use of low-cost switch networks in VPS-HPD. Moreover, both VPS-HPD and VPS-LC-HPD outperform FPS-AltMin with the same number of phase shifters and the same power consumption, since both VPS-HPD and VPS-LC-HPD can sufficiently exploit the flexibility of different phases by optimizing the phase matrix according to the CSI. The SE of VPS-LC-HPD is lower than that of VPS-HPD, but the computational complexity of VPS-LC-HPD is much lower than that of VPS-HPD.

\begin{figure}[!t]
\centering
\includegraphics[width=115mm]{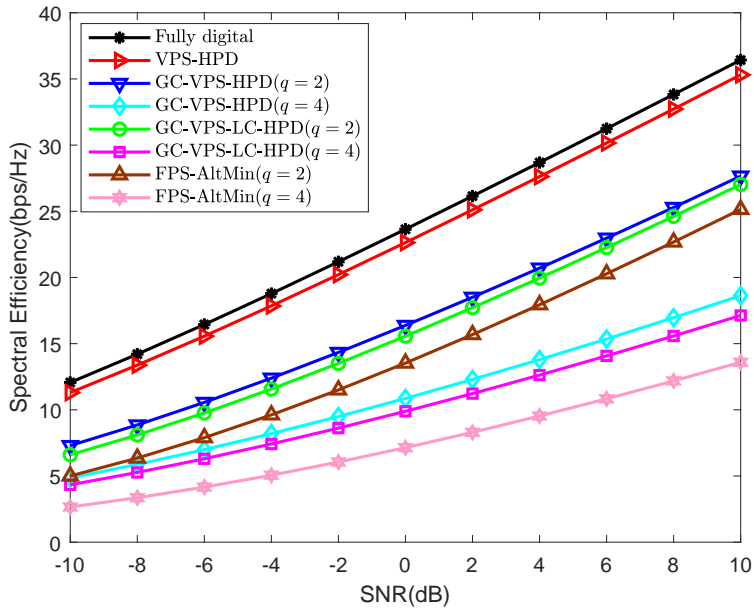}
\caption{SE comparisons of the GC-VPS architecture with the other ones.}
\label{fig:DIS3}
\end{figure}

In Fig.~\ref{fig:DIS} we focus on the VPS architecture and compare it with the others, while in Fig.~\ref{fig:DIS3} we focus on the GC-VPS and compare it with the others. When evaluating the SE performance of the FPS-AltMin scheme for the GC-VPS architecture, we fix the phases of the PSNs for GC-VPS. As shown in Fig.~\ref{fig:DIS3}, we compare the SE performance for various HPD schemes including GC-VPS-HPD, GC-VPS-LC-HPD, VPS-HPD and FPS-AltMin. It is seen that there is around $6.26~{\rm bps/Hz}$ drop in the SE performance when changing from the VPS architecture to GC-VPS with $q=2$ using the same VPS-HPD scheme. There is another $5.52~{\rm bps/Hz}$ drop in the SE performance using the same VPS-HPD scheme for the GC-VPS architecture when changing from $q=2$ to $q=4$. Therefore, the SE performance of the GC-VPS architecture will decrease if we reduce the power consumption coming from the switches. In particular, for the same GC-VPS architecture, the performance gap between the VPS-HPD scheme and the FPS-AltMin scheme gets larger when $q$ grows. If we enlarge $q$ from $2$ to $4$, the SE gap will increase from $2.85~{\rm bps/Hz}$ to $3.72~{\rm bps/Hz}$. Therefore, the freedom of phases is more important if there are fewer switches available, which verifies the effectiveness of the VPS-HPD scheme.

\begin{figure}[!t]
\centering
\includegraphics[width=115mm]{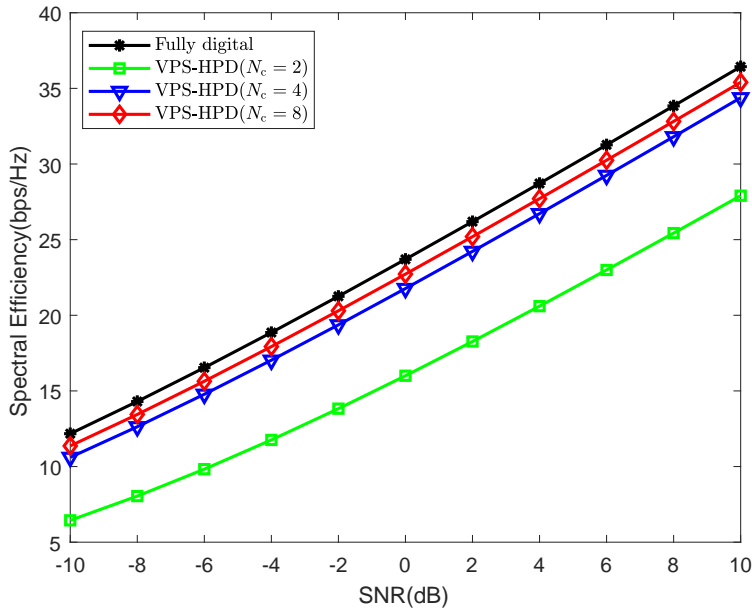}
\caption{SE comparisons of the VPS-HPD scheme with different $N_{\rm c}$.}
\label{fig:DIS2}
\end{figure}

In Fig.~\ref{fig:DIS2}, we compare the SE performance of VPS-HPD for different $N_{\rm c}$. From this figure, larger $N_{\rm c}$ leads to better SE performance. When increasing $N_{\rm c}$ from $N_{\rm c}=2$ to $N_{\rm c}=4$, around $5.77~{\rm bps/Hz}$ improvement of SE can be achieved at ${\rm SNR}=0~{\rm dB}$. But only $0.95~{\rm bps/Hz}$ improvement of SE can be achieved at ${\rm SNR}=0~{\rm dB}$, when further increasing $N_{\rm c}$ from $N_{\rm c}=4$ to $N_{\rm c}=8$. Since the SE of VPS-HPD with $N_{\rm c}=8$ is close to the performance upper bound, the SE improvement obtained by further increasing $N_{\rm c}$ may be insignificant compared with the growing power consumption of phase shifters.

\subsection{Power Consumption Comparison}\label{subsec.simu2}
In Table~\ref{Table2}, we compare the power consumption for different architectures, including the fully-connected, FPS, VPS and GC-VPS architectures. Since the same numbers of antennas and RF chains are used for different architectures, we focus on the total power consumption coming from the phase shifters and switches. The number of phase shifters used by the transmitter and receiver for FPS, VPS and GC-VPS is the same $2N_{\rm c}N_{\rm RF}=64$, while that for the fully-connected architecture is $(N_{\rm t}+N_{\rm r})N_{\rm RF}=320$. The number of switches used by the transmitter and receiver for FPS and VPS is the same $(N_{\rm t}+N_{\rm r})N_{\rm RF}N_{\rm c}=2560$, while that for GC-VPS with $q=2$ or $q=4$ is $(N_{\rm t}+N_{\rm r})N_{\rm RF}N_{\rm c}/q=1280$ or $640$, respectively. Since the power consumption of each phase shifter or each switch is $30~{\rm mW}$ or $1~{\rm mW}$~\cite{pay2019pssw}, respectively, the total power consumption of the fully-connected, FPS, VPS, GC-VPS with $q=2$ and GC-VPS with $q=4$ architectures is $9.6~{\rm W}$, $4.48~{\rm W}$, $4.48~{\rm W}$, $3.2~{\rm W}$ and $2.56~{\rm W}$, respectively.

From the above discussion, the total power consumption of the fully-connected architecture is more than twice of that of the FPS or VPS architectures, while the SE performance of the fully-connected architecture is still worse than that of VPS and only slightly better than that of FPS according to Fig.~\ref{fig:DIS}. Therefore, the effectiveness of replacing high-resolution phase shifters by the low-cost switches is verified. To further reduce the power consumption, GC-VPS with larger $q$ can be considered. Compared with the FPS or VPS architecture, the GC-VPS architecture with $q=4$ can reduce the total power consumption by almost half.


\subsection{Energy Efficiency Comparison}\label{subsec.simu3}
\begin{figure}[!t]
\centering
\includegraphics[width=115mm]{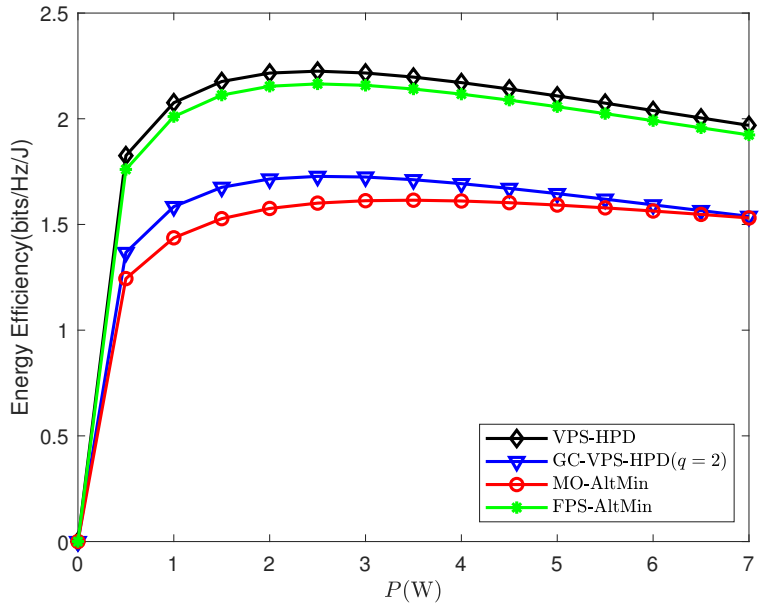}
\caption{Comparisons of energy efficiency for different architectures.}
\label{fig:EE}
\end{figure}
In this subsection, we compare the EE for different architectures including the VPS, GC-VPS, FPS and fully-connected architectures and set $q=2$. The EE is defined as the ratio of the SE over the total power consumption as
\begin{equation}\label{EE-Def}
  \eta = \frac{R}{P + N_{\rm RF}P_{\rm RF}+N_{\rm t}P_{\rm PA}+N_{\rm PS}P_{\rm PS}+N_{\rm SW}P_{\rm SW}},
\end{equation}
where $P_{\rm RF}$, $P_{\rm PS}$, $P_{\rm PA}$ and $P_{\rm SW}$ are the power of each RF chain, phase shifter, power amplifier and switch, respectively, and $P$ is the pure transmit power defined in \eqref{system model}. We set $P_{\rm RF}=100~{\rm mW}$, $P_{\rm PA}=100~{\rm mW}$, $P_{\rm PS}=30~{\rm mW}$ and $P_{\rm SW}=1~{\rm mW}$~\cite{Xianghao2016alter}. It is seen from Fig.~\ref{fig:EE} that as $P$ increases, the EE first grows and then drops. The VPS architecture achieves better EE than the other ones, since the VPS architecture adopts the mixture use of the PSNs and switch network and allows the phases of the PSNs to be variable. In particular, GC-VPS-HPD outperforms MO-AltMin. Although the SE of GC-VPS-HPD is smaller than that of MO-AltMin according to Fig.~\ref{fig:DIS} and Fig.~\ref{fig:DIS3}, the power consumption of GC-VPS-HPD is much smaller than that of MO-AltMin, which results in better EE of the former than the latter.

\subsection{Computational Complexity Comparison}\label{subsec.simu3}
The computational complexity for the VPS-HPD and VPS-LC-HPD schemes is provided in \eqref{VPS-HPD-Complexity} and \eqref{LC-VPS-HPDcomplexity}, respectively. The computational complexity of the FPS-AltMin scheme is
\begin{align}\label{FPS-AltMinComplexity}
  \mathcal{O}\Big(N_{\rm max}^{(4)}\big(N_{\rm s}N_{\rm t}  N_{\rm RF}N_{\rm c}  +N_{\rm c}N_{\rm RF}N_{\rm t}\log_2 ( N_{\rm c}N_{\rm RF}N_{\rm t})\big)\Big)
\end{align}
where $N_{\rm max}^{(4)}$ is a predefined number of iterations. The computational complexity of the MO-AltMin scheme is
\begin{equation}\label{MO-AltMinComplexity}
   \mathcal{O}\big( N_{\rm max}^{(5)} (N_{\rm t}N_{\rm RF}^2 +  \zeta )  \big)
\end{equation}
where $\zeta$ represents the number of complex-valued multiplication to obtain a solution using the Riemannian manifold
optimization and $N_{\rm max}^{(5)}$ is a predefined number of iterations.

To precisely compare the computational complexity of the VPS-HPD, VPS-LC-HPD, FPS-AltMin and MO-AltMin schemes, we measure their running time of the simulation under the same computer hardware and software. Note that these algorithms might have already converged after a fixed number of iterations. Therefore, stopping the iterations when the gap of the objective function between two adjacent iterations is smaller than a threshold, is better than stopping after a fixed number of iterations. Taking VPS-HPD as an example, we set the stop condition as $|a_n - a_{n+1}|/a_n < 0.001$, where $a_n$ is the value of the objective function at the $n$th iteration. Then the time of convergence for the VPS-HPD, VPS-LC-HPD, FPS-AltMin and MO-AltMin schemes is $13.90~{\rm s}$, $0.089~{\rm s}$, $0.065~{\rm s}$ and $9.91~{\rm s}$, respectively, implying that the computational complexity of VPS-LC-HPD is much lower than that of VPS-HPD.


\section{Conclusions}\label{sec.conclusion}
In this paper, we have proposed the VPS architecture, whose phases are variable and can be optimized according to the CSI subject to the hardware constraints. Based on the VPS architecture, we have proposed the VPS-HPD scheme by alternately optimizing the analog precoder and the digital precoder. To reduce the computational complexity, we have proposed the VPS-LC-HPD scheme that does not need Riemannian manifold optimization and
exhaustive search. To reduce the hardware complexity in terms of large number of switches, we have considered the GC-VPS architecture and proposed a HPD scheme that solves multiple independent subproblems by the VPS-HPD or VPS-LC-HPD scheme. In the future, the work will be extended to the multiuser scenario, with our focus on the low-complexity channel estimation algorithms for the VPS and GC-VPS architectures.


\bibliographystyle{IEEEtran}
\bibliography{IEEEabrv,IEEEexample}

\end{document}